\documentclass[showpacs,aps,pre]{revtex4}
\usepackage{amsmath,amssymb}
\usepackage{graphicx}
\usepackage{clrscode}
\include{epsf}

\def\(({\left(}
\def\)){\right)}                       
\def\[[{\left[}
\def\]]{\right]}

\newcommand{\be}{\begin{equation}}
\newcommand{\ee}{\end{equation}}
\newcommand{\bea}{\begin{eqnarray}}
\newcommand{\eea}{\end{eqnarray}}

\begin{document}

\title{Constraint satisfaction problems with isolated solutions are hard}

\author{Lenka Zdeborov\'a}
\affiliation{ Universit\'e Paris-Sud, LPTMS, UMR8626,  B\^{a}t.~100, Universit\'e
Paris-Sud 91405 Orsay cedex}
\affiliation{CNRS, LPTMS, UMR8626, B\^{a}t.~100, Universit\'e Paris-Sud 91405 Orsay cedex}
\affiliation{Theoretical Division and Center for Nonlinear Studies, Los Alamos National Laboratory, Los Alamos, NM 87545}

\author{Marc M\'ezard}
\affiliation{ Universit\'e Paris-Sud, LPTMS, UMR8626,  B\^{a}t.~100, Universit\'e
Paris-Sud 91405 Orsay cedex}
\affiliation{CNRS, LPTMS, UMR8626, B\^{a}t.~100, Universit\'e Paris-Sud 91405 Orsay cedex}

\begin{abstract}
  We study the phase diagram and the algorithmic hardness of the random
  `locked' constraint satisfaction problems, and compare them to the commonly studied 'non-locked'
  problems like satisfiability of boolean formulas or graph coloring. The special property of the
  locked problems is that clusters of solutions are {\it isolated} points. This simplifies
  significantly the determination of the phase diagram, which makes the locked 
  problems particularly appealing from the mathematical point of view. On the
  other hand we show empirically that the clustered phase of these
  problems is extremely hard from the algorithmic point of view:  the best
  known algorithms all fail to find solutions. Our results suggest that the
  easy/hard transition (for currently known algorithms) in the locked problems
  coincides with the clustering transition. These should thus be regarded as
  new benchmarks of really hard constraint satisfaction problems.
\end{abstract}

\date{\today}

\maketitle

\section{Introduction}
Constraint satisfaction problems (CSPs) play a crucial role in theoretical and
applied computer science. Their wide range of applicability arises from their
very general nature: given a set of $N$ discrete variables subject to $M$
constraints, a CSP consists in deciding whether there exists an assignment
of variables which satisfies simultaneously all the constraints. When such an
assignment exists we call it a solution and aim at finding it. One of the most
important questions about a CSP is how hard it is to
find a solution or prove that there is none. Many of the CSPs belong to the
class of NP-complete problems \cite{Cook71,GareyJohnson79}. This basically
means that, if P$\neq$NP, there is no algorithm able to solve the worst case
instances of the problem in a polynomial time.
Next to the question of the {\it worst case} computational complexity arises
the less explored question of  {\it typical case}
complexity. A pivotal step in understanding the typical case complexity is the
study of {\it random} CSPs where each constraint
involves a finite number of variables. Pioneering work on this subject
\cite{CheesemanKanefsky91,MitchellSelman92} discovered that many problems are
empirically harder close to the so-called satisfiability phase transition. This is a
phase transition appearing at a critical constraint density $\alpha_s$ such
that for $M/N=\alpha<\alpha_s$ almost every large instance of the problem has
at least one solutions, and for $\alpha>\alpha_s$ almost all large instances
have no solution.

Studies of phase transitions such as the one occuring in the satisfiability problem are natural for
statistical physicists. Indeed the methods developed to study frustrated
disordered systems like glasses and spin glasses \cite{MezardParisi87b} have turned
out to be very fruitful in the study of several CSPs. In particular they allow
some structural studies which aim at understanding how the difficulty of a problem
is related to the geometrical organization of its solutions. Several other phase
transitions were described in this context. The most important one is probably
the clustering transition \cite{MezardParisi02,BiroliMonasson00}, known as the
dynamical glass transition in the mean field theory of glasses. It was computed that in the region where the density of constraints is below the satisfiability threshold there exists a phase where the space
of solutions splits into ergodically separated groups -- clusters. Another
important property of the clusters concerns the freezing of the variables. A
variable is frozen in a cluster if it takes the same value in all the
solutions of this cluster. It has been conjectured that the clustering
 \cite{MezardZecchina02} and the freezing of variables 
 \cite{ZdeborovaKrzakala07} are two ingredients which contribute to
make a random CSP hard. But the predictions for the easy/hard transition in a general random CSP are still not fully quantitative. The present work
provides further insight into this subject.

In this paper we present a detailed study of the {\it locked} CSPs, introduced recently in
\cite{ZdeborovaMezard08}. The special property of the locked problems is that
clusters are point-like: every cluster contains only one solution. Therefore, as soon as the
system is in a clustered phase, all the variables are frozen in each cluster. The clustering and
the freezing phase transitions occur simultaneously.
Consequently the organization of the space of solutions is much simpler than
in the commonly studied K-satisfiability or graph coloring
\cite{MezardParisi02,MuletPagnani02,KrzakalaMontanari06,ZdeborovaKrzakala07,MontanariRicci08}.
But at the same time, and unlike in the K-satisfiability or graph coloring problems, the whole clustered phase is extremely hard for all existing
algorithm and the clustering/freezing threshold seems to coincide
very precisely with the onset of this hardness.

The interest in the locked problems is thus twofold: 
\begin{itemize}
\item[(a)]{{\bf Locked problems are very simple:} As the clusters of solutions are
    point-like many of the quantities of interest can be computed using
    simpler tools than in the canonical K-satisfiability problem. This is in
    particular interesting from the mathematical point of view, because several of their
    properties become accessible to rigorous proofs. From a broader point of
    view the locked problems should be useful as simple models of glass
    forming liquids because their phase diagram can be studied without any need to
    introduce the complicated scheme of `replica symmetry
    breaking' \cite{MezardParisi87b}.}
    \item[(b)]{{\bf Locked problems are very hard:} From the algorithmic point of
        view the whole clustered phase of the locked problems is extremely hard,
        none of the known algorithms is able to find solutions efficiently.
        This suggests to use locked CSPs as hard benchmarks. At
        the same time one may hope that the performance of some algorithms
        will be simpler to analyze when they are applied to the locked problems, compared to the general case.}
\end{itemize} 

This paper is organized as follows: In section~\ref{sec:def} we define the
random occupation problems and the random locked occupation problems (LOPs) on
which we will illustrate our main findings. In section~\ref{sec:cav}
we write the equations needed to describe the phase diagram of the occupation
problems, using well known tools from statistical physics and probability
theory. In section~\ref{sec:phase} we summarize the basic properties of the
phase diagram in general random CSPs and then discuss in detail the situation
in the locked problems. We also discuss the class of so-called balanced LOPs
which are even simpler from the mathematical point of view. Finally
section~\ref{sec:alg} shows our findings about algorithmic performance in the
occupation problems:  empirical data using the best known random CSP solver
-- belief propagation reinforcement -- indicates that the clustering
threshold is close to the boundary between the easy and hard regions. We
analyze also the non-locked occupation problems for comparison. A short summary
of the results and perspectives conclude the paper in section~\ref{sec:con}.

\section{Definitions}
\label{sec:def}

\subsection{Locked occupation problems}

We shall study  a broad class of problems called `{\it occupation problems}'. An occupation
problem involves  $N$
binary variables $s_i\in\{0,1\}$  ($s_i=0$ is referred to as ``site $i$ is
empty'', and $s_i=1$ is ``occupied'')
and $M$ constraints, indexed by $b\in\{1,\dots ,M\}$.
Each constraint $b$ involves $K_b$ distinct variables, and is defined by a 
`constraint word' $A^{b}$ with 
 $K_b+1$ bits, which we write as $A^{b}=A_0^{b}A_1^{b}\dots A_{K_b}^{b}$, where $A_i^{b}\in\{0,1\}$.
We denote by  $\partial b$ the indices of all variables involved in 
the constraint $b$. The constraint $b$
is satisfied if and only if the sum $r=\sum_{i\in \partial b} s_i$ of all  its variables
is such   that $A_r^{b}=1$. In other words, in order for
constraint $b$ to be satisfied, one needs  that the number of occupied sites, $r$, in its
neighborhood, must be such that $A_r^{b}=1$ (this unified notation for the
occupation problems was introduced in \cite{Mora07}). 

{\bf Definition}: An occupation problem is {\it locked} if and only if:
\begin{itemize}
\item[(a)]{For every constraint $b\in\{1,\dots,M\}$, the vector $A^{b}$ is such that,
for all $i=0,\dots,K-1$: $A_i^{b}A_{i+1}^{b}=0$ .} 
\item[(b)]{Every variable appears in at least two different constraints.}
\end{itemize}

\begin{figure}[!ht]
    \resizebox{0.35\linewidth}{!}{\includegraphics{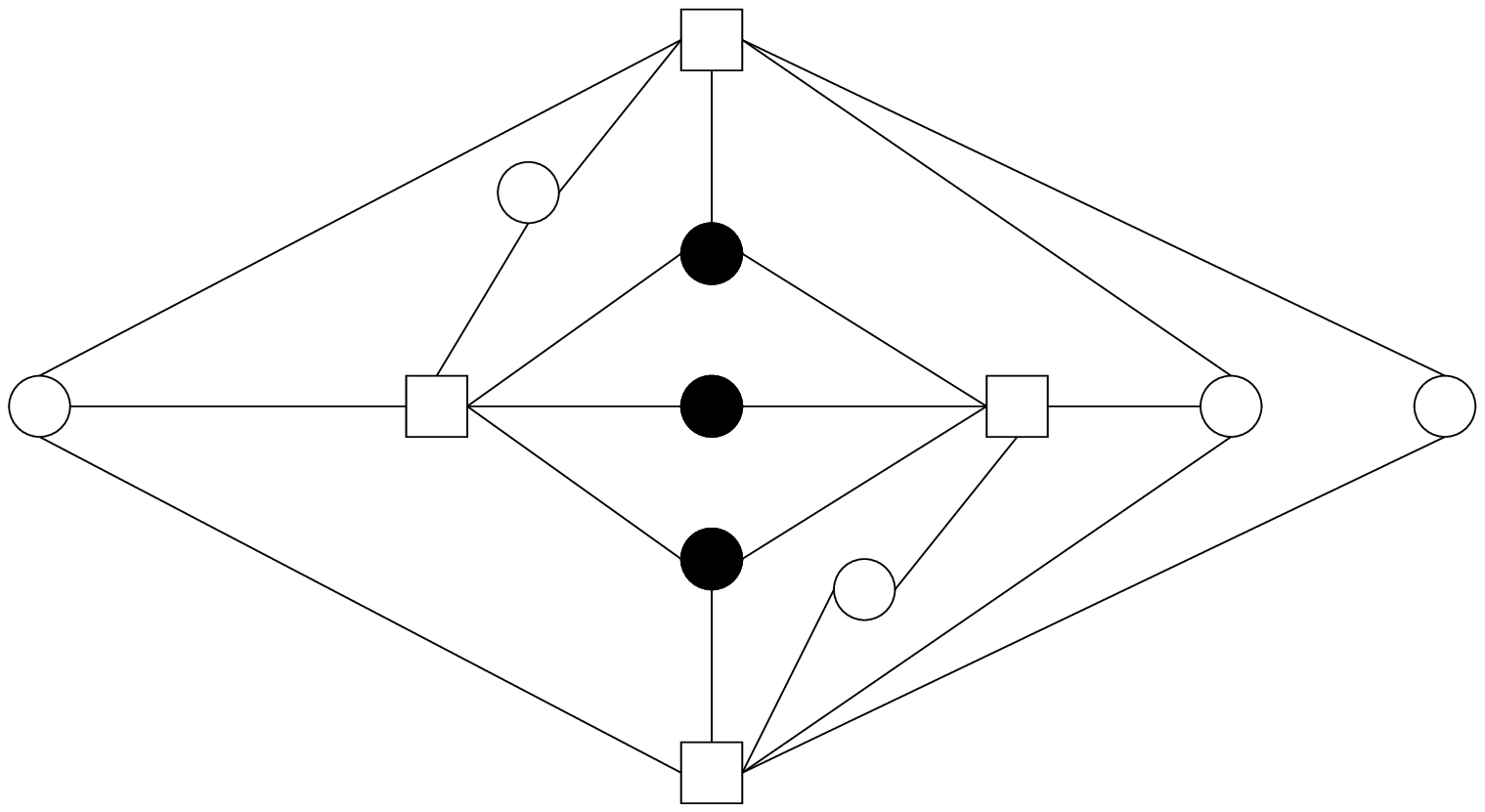}}
    \resizebox{0.1\linewidth}{!}{\hspace{1.5cm}}
    \resizebox{0.35\linewidth}{!}{\includegraphics{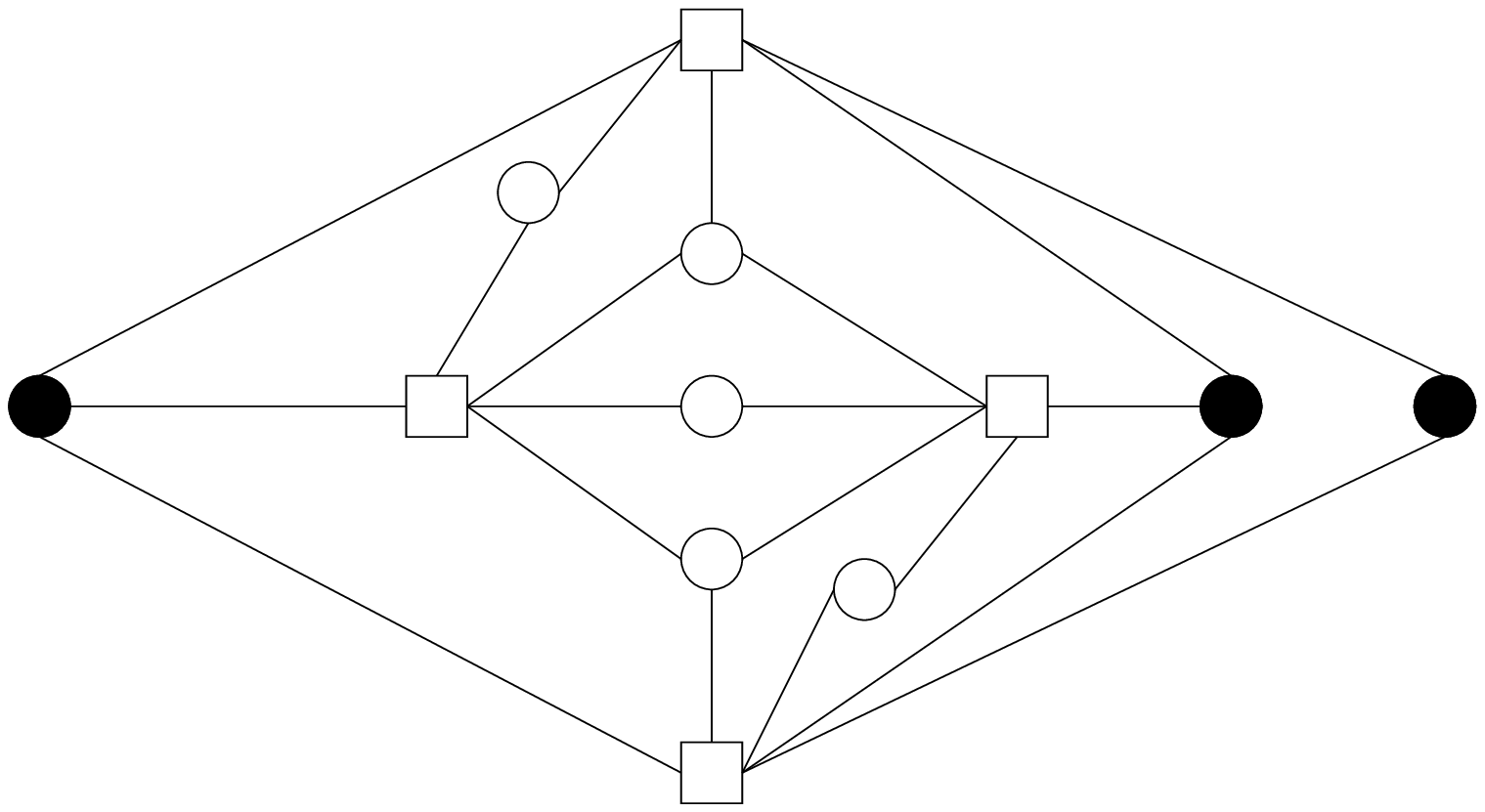}}
  \caption{\label{Fig:example} A factor graph representation of an instance of the 1-or-3-in-5 SAT ($A=010100$). The squares are the constraints. Full/empty circles are occupied/empty sites. The two parts show two examples of satisfying assignments -- ``solutions'' -- of this instance. As there are no leaves (each variable belongs to at least two constraints) and $A$ satisfies $A_iA_{i+1}=0$ for all $i=0,\dots,4$, this instance is locked.}
\end{figure}

In this paper, we shall study only `constraint-regular' problems in which all of the constraints
are described by the same constraint word: for all $b\in\{1,\dots,M\}$, $K_b=K$ and $A^{b}=A$.
Furthermore, in order to focus onto difficult cases, we shall only consider the 
occupation problems where neither the totally empty nor the totally occupied
configurations are solution, i.e. we keep to the cases where $A_0=A_K=0$.
It is convenient to use the factor graph description of a problem
\cite{KschischangFrey01,MezardMontanari07}, where sites and constraints are vertices, and an edge
connects a constraint $a$ to a site $i$ whenever $i$ appears in constraint $a$
(see Fig.~\ref{Fig:example}). An instance of a constraint-regular occupation
model is fully described by its factor graph (where all constraint vertices
have degree~$K$) and the $K+1$ component vector $A$. The locked problems are thus characterized
by the facts that (i) there are no consecutive `1' in the word $A=A_0A_1\dots A_{K}$, and (ii) their 
factor graph has no leaves.

Well-studied examples of occupation problems include: 
\begin{itemize}
    \item{Ising anti-ferromagnet: $A=010$ }
    \item{Odd parity checks (anti-ferromagnetic $K$-spin model, with $K$ even): $A=01010\dots1010$ }
    \item{Positive 1-in-K SAT (exact cover): $A=0100\dots00$ \cite{RaymondSportiello07}}
    \item{Perfect matching in $K$-regular graphs: each variable belongs to two constraints and $A=01000\dots00$ \cite{ZdeborovaMezard06}}
    \item{Bicoloring (positive NAE-SAT): $A=0111\dots110$ \cite{AchlioptasMoore06,CastellaniNapolano03,DallAstaRamezanpour08}} 
    \item{Circuits going through all the points: $A=001000\dots00$ \cite{MarinariSemerjian06}}
\end{itemize}
All these examples, except the bicoloring, are locked on graphs without leaves.

For the occupation problems which have not been studied previously,
 we will use names derived in the following way: 
$A=010100$ is the 1-or-3-in-5 SAT, $A=010010$ is the 1-or-4-in-5 SAT, etc.

From the computational complexity point of view, Schaefer's theorem
\cite{Schaefer78} implies that most of the occupation problems are NP-complete. The exceptions are
 the parity checks, which amount to linear systems of equations on $GF(2)$, and some of the cases where the variables
have degree 2, such as for instance the perfect matching.

The crucial property of the locked occupation problems is that in order to go from one solution to another one must flip at least a closed loop of variables. This property can be used to generalize the definition of a locked problems to a much wider class of constraint satisfaction problems than the occupation problems, and
in particular the variables do not need to be binary. 
Some examples of locked problems which are not occupation problems
 are the XOR-SAT (p-spin) problem on factor graphs without leaves \cite{MezardRicci03}, 
or all the uniquely extensible models \cite{ConnamacherMolloy04}.  

\subsection{Ensembles of random occupation problems}

We shall study some random ensembles of locked occupation problems, in which the factor graph is
chosen from some ensemble of random bipartite graphs. We consider constraint-regular occupation problems
where each constraint involves $K$ variables, and is characterized by the constraint word $A$.
An ensemble is characterized via a probability distribution $Q(l)$.
To create an instance of the random occupation problem with $N$ variables, we
draw $N$ independent random numbers $l_i$ from the distribution $Q(l)$, with
the additional constraint that $\sum_{i=1}^N l_i/K = M$ is an integer. The
factor graph that characterizes an
instance 
is then chosen uniformly at random from all the possible graphs
with $N$ variables, and $M$ constraints, such that, for all $i=1\dots N$, the variable $i$
is connected to $l_i$ constraints.

In this paper we will consider mainly two degree distributions:
\begin{itemize}
   \item{Regular: $Q(l)=\delta_{l,L}$, in which all the variables take part in $L$ clauses.}
   \item{Truncated Poissonian: 
\be 
  Q(0)=Q(1)=0\, , \quad \quad   Q(l)= \frac{1}{1 - (1 + c)\, e^{-c}} \frac{e^{-c} c^l}{l!} \quad {\rm for} \quad  l\ge 2 \label{Poiss}
\ee 
where $c\ge 0$. The average ``connectivity'' (variable degree) is then 
\be 
\overline l = c\, \frac{1 - e^{-c}}{1 - (1 + c) \, e^{-c}}\ . \label{l_aver}
\ee
In the cavity method one also needs the excess degree distribution $q(l)$, defined as the distribution
of the number of neighbors on one side of an edge chosen uniformly at random:
\be
       q(0) = 0\, ,    \quad \quad  q(l)=\frac{1}{e^c-1} \frac{c^k}{k!}
\label{excess}
\, .
\ee
}
\end{itemize} 
 We
shall be interested in the properties of large instance, i.e. in the `thermodynamic limit' where 
 one sends $N\to \infty$ and  $M\to\infty$, keeping $K$ and $Q(l)$ fixed; this results in a 
 fixed density of constraints
 $M/N=\overline l/K$. Our main results are easily generalizable to any degree distribution $Q(l)$ which has a finite second moment. For every such distribution, a typical factor graph is locally tree-like: the shortest loop going through a typical variable has a length which scales as $\log{N}$.  
The crucial property of the locked occupation problems is that, in order to go from one
solution of the problem to another solution, one must flip at least one closed loop of variables. On
the random locally tree-like factor graphs this means that at least $\log{N}$
variables need to be changed.

\section{The solution of random occupation problems}
\label{sec:cav}

The cavity method \cite{MezardParisi01} is nowadays the standard tool to
compute the phase diagram of random locally tree-like constraint satisfaction
problems. Depending on the structure of the space of solutions of the problem,
different versions (levels of the replica symmetry breaking) of the method are
needed. In this section we state the cavity equations for the occupation
problems. For a detailed derivation and discussion of the method see
\cite{MezardParisi01,MezardMontanari07}.

We index the variables by $i,j,k,\dots$ going from $1$ to $N$, and the
constraints by $a,b,c,\dots$ going from $1$ to $M$. The energy of the
occupation problems then reads 
\be 
H(\{s\}) = \sum_{a=1}^M
\delta_{A_{\sum_{j\in \partial a} s_j},0}\, , 
\ee 
In this paper we shall study only the instances where solutions (ground
states of zero energy) exist, and we shall focus on the uniform measure over all
solutions.

\subsection{The replica symmetric solution}

\begin{figure}[!ht]
  \resizebox{4cm}{!}{\includegraphics{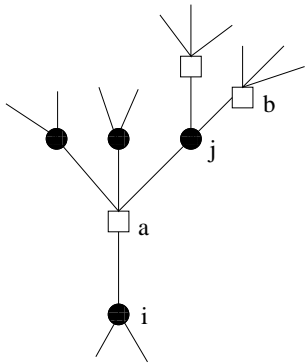}}
  \caption{\label{fig_cav}Part of the factor graph to illustrate the meaning of indices in the belief propagation equations (\ref{BP1}-\ref{BP2}).}
\end{figure}

The replica symmetric version of the cavity method is also known under the name  belief propagation \cite{Pearl82,KschischangFrey01,MezardMontanari07}. It exploits  the local tree-like property of the factor graph, assuming that correlations decay fast enough. The basic quantities
used in this approach are messages. We define $\psi_{s_i}^{a\to i}$ as the probability that the constraint $a$ is satisfied, conditioned to the fact that the value of  variable $i$ is $s_i$. Similarly,  $\chi_{s_j}^{j\to a}$ is the probability that the variable $j$ takes value $s_j$ conditioned to the fact that the constraint $a$ has been removed from the graph. The messages then satisfy the belief propagation (BP) equations
\begin{subequations}
\label{eq:BP_locked}
\bea
     \psi_{s_i}^{a\to i} &=& \frac{1}{Z^{a\to i}} \sum_{\{s_j\}} \delta_{A_{s_i+\sum_{j} s_j},1}\prod_{j\in \partial a-i} \chi_{s_j}^{j\to a} \, , \label{BP1} \\
     \chi_{s_j}^{j\to a} &=& \frac{1}{Z^{j\to a}} \prod_{b\in \partial j-a} \psi_{s_j}^{b\to j}\, , \label{BP2} 
\eea
\end{subequations}
where $Z^{a\to i}$ and $Z^{j\to a}$ are normalization constants. Fig.~\ref{fig_cav} shows the corresponding part of the factor graph. The marginal probabilities (``beliefs'') are then expressed as
\be
    \chi_{s_i}^{i} = \frac{1}{Z^{i}} \prod_{a\in \partial i} \psi_{s_i}^{a\to i}\, , \label{marginal}
\ee

The  replica symmetric entropy (logarithm of the number of solutions, divided by the system size) then reads
\be
   s=\frac{1}{N}\sum_a \log{(Z^{a})} -  \frac{1}{N} \sum_i (l_i-1)  \log{(Z^i)}\, . \label{eq:entropy_l}
\ee
where
\begin{subequations}
\label{eq:Z_locked}
\bea
     Z^{a}&=& \sum_{\{s_i\}} \delta_{A_{\sum_i s_i},1} \prod_{i\in \partial a} \left( \prod_{b\in \partial i-a} \psi_{s_i}^{b\to i} \right)\, , \label{Za}\\
     Z^i&=&  \prod_{a\in \partial i} \psi_{0}^{a\to i}+ \prod_{a\in \partial i} \psi_{1}^{a\to i}\, , \label{Zi}
\eea 
\end{subequations}
 are the exponentials of the entropy shifts when the node $a$ and its neighbors (resp. the node $i$) is added. 

 When one considers an ensemble of random graphs, the probability distribution
 of the messages can be found via the population dynamics technique
 \cite{MezardParisi01}. Moreover, on the regular graph ensemble or for some
 of the balanced problems (see Sec.~\ref{sec:bal}) the solution is {\it
   factorized}. In the factorized solution the messages $\chi^{i\to a}$,
 $\psi^{a \to i}$ are independent of the edge $(ia)$ and the replica symmetric solution can thus be found analytically.

For instance in the regular graph ensemble where each variable is present in $L$ constraints the factorized solution is  
\begin{subequations}
\label{eq:locked_reg}
\bea
 \psi_0&=& \frac{1}{Z^{\rm reg}}  \sum_{r=0}^{K-1} \delta_{A_r,1} {K-1 \choose r} \psi_1^{(L-1)r} \psi_0^{(L-1)(K-1-r)}\ , \label{RS1}\\
 \psi_1&=& \frac{1}{Z^{\rm reg}}  \sum_{r=0}^{K-1} \delta_{A_{r+1},1}  {K-1 \choose r} \psi_1^{(L-1)r} \psi_0^{(L-1)(K-1-r)} \, ,\label{RS2}
\eea 
\end{subequations}
where the normalization $Z^{\rm reg}$ is fixed by the condition $\psi_0+\psi_1=1$.
Given the solution $\psi_0$, $\psi_1$ of (\ref{RS1}-\ref{RS2}), the entropy reads
\be
  s_{\rm reg}= \frac{L}{K} \log{\left[ \sum_{r=0}^K \delta_{A_r,1}  {K \choose r} \psi_1^{(L-1)r} \psi_0^{(L-1)(K-r)}\right]} -
  (L-1) \log{\left[\psi_{0}^{L}+  \psi_{1}^{L}\right]}\, .
\ee

\subsection{Reconstruction on trees}

Treating the locally tree-like random graph as a tree fails if long range
correlations are present in the system. More precisely
\cite{MezardMontanari06,KrzakalaMontanari06} the replica symmetric assumption
is correct if and only if the so-called point-to-set correlations do decay to
zero. The decay of these correlations is closely related \cite{MezardMontanari06} to the problem of
reconstruction on trees \cite{Mossel01} which we explain and analyze
in this section.

The reconstruction on trees is defined as follows: First construct a tree of
$d$ generations having the same connectivity properties as a finite
neighborhood of a random variable in the random factor graph. Assign the root
a random value, further assign values iteratively on the descendants uniformly
at random but in such a way that the constraints are satisfied. Subsequently
forget the assignment everywhere but on the leaves of the tree. The
reconstruction on the tree is possible if and only if the information left in
the values of the leaves about the value of the root does not go to zero as
the size of the tree grows, $d \to \infty$. The replica symmetric assumption
is correct if and only if the reconstruction is not possible, in other words
if there is no correlation between the root (point) and the leaves (set). Typically, when the average
connectivity of variables is small the reconstruction is not possible and when
the connectivity is large the reconstruction is possible. The threshold
connectivity is called the reconstructibility threshold, or the clustering
transition. The clustering is then defined as a minimal decomposition of the
space of solutions such that within the components (clusters) the point-to-set
correlations do decay to zero \cite{KrzakalaMontanari06}.

It was shown in \cite{MezardMontanari06} that the analysis of the
reconstruction on trees is equivalent to the solution of the one-step replica
symmetry breaking (1RSB) equations at the value of the Parisi parameter $m=1$
\cite{MezardParisi01}.

Instead of the general form of the 1RSB equations at $m=1$ (see e.g.~\cite{Zdeborova08}), we shall only
discuss here a simpler form called the {\it naive reconstruction} in
\cite{Semerjian07}. In general the naive reconstruction gives only an upper
bound on the reconstructibility threshold, but in the locked problems it gives
in fact the full information. The naive reconstruction consists
in computing the probability that the value of the root is {\it uniquely} implied by
the leaves (boundary conditions). Here we give the equations
only for regular graph ensembles with variables of connectivity $L$, where the factorized replica symmetric solution (\ref{eq:locked_reg}) holds. 
Define $\mu_1$ (resp. $\mu_0$) as the probability that a variable which in the broadcasting had value $1$ (resp. $0$) is uniquely determined by the boundary conditions.
One has:
\begin{subequations}
\label{eq:mus}
\bea
   \mu_1 &=& \frac{1}{\psi_1 Z^{\rm reg}}\sum_{r=0}^k 
\delta_{A_{r+1},1}\delta_{A_r,0 }\,   g_k(r) \sum_{s=0}^{s_1} {r \choose s} \left[1-(1-\mu_0)^{l}\right]^{k-r} \left[1-(1-\mu_1)^{l}\right]^{r-s} (1-\mu_1)^{ls} \, ,  \label{eq:mu1}\\
    \mu_0 &=& \frac{1}{\psi_0 Z^{\rm reg}} \sum_{r=0}^k 
\delta_{A_{r+1},0}\delta_{A_r,1 } 
 \, g_k(r) \sum_{s=0}^{s_0} {k-r \choose s} \left[1-(1-\mu_1)^{l}\right]^{r}  \left[1-(1-\mu_0)^{l}\right]^{k-r-s} (1-\mu_0)^{ls}\, ,\label{eq:mu0} 
\eea
\end{subequations}
where $l=L-1$, $k=K-1$, and $g_k(r)={k \choose r} (\psi_1)^{lr}  (\psi_0)^{l(k-r)} $ . The indices $s_1,s_0$ in the second sum of both equations are the largest possible but such that $s_1\le r$, $s_0\le K-1-r$, and $\sum_{s=0}^{s_1}A_{r-s}=0$, $\sum_{s=0}^{s_0}A_{r+1+s}=0$.
The values $\psi_0$, $\psi_1$ are the fixed point of eqs.~(\ref{RS1}-\ref{RS2}), and $Z^{\rm reg}$ is the corresponding normalization.

These lengthy equations have in fact a simple meaning. The first sum is over
the possible numbers $r$ of occupied variables on the descendants in the
broadcasting. The sum over $s$ is over the number of variables which were not
implied by at least one constraint but the configuration of implied variables
nevertheless implies the outcoming value. The term $1-(1-\mu)^l$ is the probability
that at least one constraint implies the variable, $(1-\mu)^l$ is the
probability that none of the constraints implies the variable.

\subsection{Survey Propagation}

Survey propagation is a special form of the 1RSB equations corresponding to the value of the Parisi parameter $m=0$ \cite{MezardZecchina02}. The main assumption of the 1RSB approach is that the space of solutions splits into clusters (pure states). To each cluster corresponds one fixed point of BP equations. Survey propagation are then iterative equations for the following probabilities (surveys)
\begin{subequations}
\label{eq:probs}
\bea
   {\rm Prob}(\chi_1^{i \to a}=1,\chi_0^{i \to a}=0) = p_1^{i \to a},   & \quad \quad & {\rm Prob}(\psi_1^{a\to i}=1,\psi_0^{a\to i}=0) = q_1^{a \to i}, \\ 
   {\rm Prob}(\chi_1^{i \to a}=0,\chi_0^{i \to a}=1) = p_0^{i \to a},   & \quad \quad & {\rm Prob}(\psi_1^{a\to i}=0,\psi_0^{a\to i}=1) = q_0^{a \to i}, \\
   p_*^{i\to a} = 1- p_1^{i \to a} - p_0^{i \to a},  & \quad \quad & q_*^{a \to i}= 1- q_1^{a \to i} - q_0^{a \to i},
\eea
\end{subequations}
where $q_{1/0}^{a\to i}$ is probability over clusters that clause $a$ is satisfied only if variable $i$ takes value $1/0$, $q_*^{a\to i}$ is then the probability that clause $a$ can be satisfied by both values $1$ and $0$, similarly $p_{1/0}^{i\to a}$ is probability that variable $i$ have to take value $1/0$ if the clause $a$ is not present, $p_{*}^{i\to a}$ is probability that the variable $i$ can take both values $1$ and $0$ when clause $a$ is not present. The survey propagation equations are then written in two steps, first the update of $p$'s knowing $q$'s
\begin{subequations}
\label{eq:SP1}
\bea
p_1^{j\to a} &=& \frac{1}{{\cal N}^{j\to  a}} \left[\prod_{b\in \partial j-a} (q_1^{b\to j} + q_*^{b\to j} ) - \prod_{b\in \partial j-a} q_*^{b\to j} \right], \\
p_0^{j\to a} &=& \frac{1}{{\cal N}^{j\to a}} \left[\prod_{b\in \partial j-a} (q_0^{b\to j} + q_*^{b\to j} ) - \prod_{b\in \partial j-a} q_*^{b\to j} \right], \\
  p_*^{j\to a} &=& \frac{1}{{\cal N}^{j\to a}} \prod_{b\in \partial j-a} q_*^{b\to j},
\eea
\end{subequations}
second the update of $q$'s knowing $p$'s
\begin{subequations}
\label{eq:SP2}
\bea
     q_s^{a\to i} &=& \frac{1}{{\cal N}^{a\to i}}  \left[ \sum_{\{r_j\}} C_s(\{r_j\}) \prod_{j\in \partial a-i} p_{r_j}^{j\to a}  \right]\ . 
\eea 
\end{subequations}
Here ${\cal N}^{j\to a} $ and ${\cal N}^{a\to i}$ are normalization constants, the
indices $s$ and $r_j$ are in 
$\{1,0,*\}$. The function $C_1$/$C_0$ (resp. $C_*$) takes
values $1$ if and only if the incoming set of $\{r_j\}$ forces the variable
$i$ to be occupied/empty (resp. let the variable $i$ free), in all other cases
the $C$'s are zero. More specifically, let us call $n_1,n_0,n_*$ the number of indices
$1,0,*$ in the set $\{r_j\}$ then
\begin{itemize}
\item{$C_1=1$ if and only if $A_{n_1+n_*+1}=1$ and $A_{n_1+n}=0$ for all $n=0\dots n_*$;} 
\item{$C_{0}=1$ if and only if $A_{n_1}=1$ and $A_{n_1+1+n}=0$ for all $n=0\dots n_*$; }
\item{$C_*=1$ if and only if there exists $m,n\in\{0\dots n_*\}$ such that $A_{n_1+n}=A_{n_1+m+1}=1$.}
\end{itemize}

\subsection{The first and the second moment}
In  this section we give the formulas for the first and second
moment method in general occupation problems.
This allows for a direct probabilistic study of the  balanced locked occupation
problems introduced below in Sec.~\ref{sec:bal}.

For a given instance (or factor graph), $G$, define as ${\cal N}_G$ the number
of solutions. The first moment is the average of ${\cal N}_G$ over the graph
ensemble, which can also be written as:  
\be 
\langle {\cal N}_G \rangle =
\sum_{\{\sigma\}} {\rm Prob}\left(\{\sigma\}\, {\rm is}\, {\rm SAT}\right)\, .
\ee 
The `annealed entropy' is then defined as $ s_{\rm ann} \equiv
\log{\langle {\cal N}_G \rangle}/N $. It is an upper bound on the quenched
entropy, $\langle \log{ {\cal N}_G} \rangle/N$. In order to compute the first
moment we divide variables into groups according to their connectivity and in
each group we choose a fraction of occupied variables. The number of ways in which
this can be done  is then multiplied by the probability that such a configuration
satisfies simultaneously all the constraints. After some algebra
\cite{Zdeborova08} we obtain the entropy of solutions with a fraction $0\le
t\le 1$ of occupied variables:
\be 
 s_{\rm ann}(t) = \sum_l Q(l)
\log{[1+u(t)^l]} + \frac{\overline l}{K} \log{\Big[\sum_{r=0}^K \delta_{A_r,1}
  {K \choose r} \left(\frac{t}{u(t)}\right)^r (1-t)^{K-r} \Big]}\,
, \label{1st_d} 
\ee
 where $u(t)$ is the inverse of 
\be t=\frac{1}{\overline l}
\sum_l l\, Q(l) \frac{u^l}{1+u^l} \, . \label{saddle_u} 
\ee
 The annealed
entropy is then $ s_{\rm ann} = {\rm max_t} s_{\rm ann}(t)$.

The second moment of the number of solutions is defined as 
\be
     \langle {\cal N}^2_G \rangle =  \sum_{\{\sigma_1\},\{\sigma_2\}} {\rm Prob}\left(\{\sigma_1\}\,  {\rm and} \,  \{\sigma_2\} \,  {\rm are} \,  {\rm both} \, {\rm SAT}\right) \, .  \label{2nd}
\ee
The second moment entropy is then defined as $s_{\rm 2nd}\equiv \log{ \langle {\cal N}^2_G \rangle }/N$. The Chebyshev's inequality gives then a lower bound on the satisfiability threshold via
\be
      {\rm Prob}({\cal N}_G>0) \ge   \frac{ \langle {\cal N}_G \rangle^2 }{ \langle {\cal N}^2_G \rangle}  \, .\label{Cheb}
\ee
The second moment is computed in a similar manner as in \cite{AchlioptasMoore06,AchlioptasNaor05}. First we fix that in a fraction $t_{x}$ of nodes the variable is occupied in both the solutions $\sigma_1,\sigma_2$ in (\ref{2nd}). In a fraction $t_{y}$ the variable is occupied in $\sigma_1$ and empty in $\sigma_2$ and the other way round for $t_{z}$. We sum over all possible realizations of $0\le t_{x}, t_{y}, t_{z}$ such that $\sum_{w=x,y,z} t_{w}\le 1$. This is multiplied by the probability that the two configurations $\sigma_1, \sigma_2$ both satisfy all the constraints. After some algebra we obtain \cite{Zdeborova08}:
\be
   s_{\rm 2nd}(t_x,t_y,t_z) = \frac{\overline l}{K} \log{p_A(t_x,t_y,t_z)}+ \sum_l Q(l) \log{\left\{1+ \sum_{w\in \{x,y,z\}} [u_w(t_x,t_y,t_z)]^l\right\}}  \, , \label{2nd_d}
\ee
where $u_w(t_x,t_y,t_z)$, $w\in \{x,y,z\}$, are obtained by inverting the three equations: 
\be
 t_w = \frac{1}{\overline l} \sum_l l\, Q(l) \frac{u_w^l}{1+ u_x^l +u_y^l+u_z^l}\, ,  \quad \quad  w=x,y,z \, , \label{saddle_tw}
\ee
and the function $p_A(t_x,t_y,t_z)$ is defined as
\bea
   p_A(t_x,t_y,t_z) &=& \sum_{r_1,r_2=0}^K  \delta_{A_{r_1}A_{r_2},1}  \sum_{s=\max{(0,r_1+r_2-K)}}^{\min{(r_1,r_2)}} {K \choose (r_1-s) (r_2-s) \, s} \left( \frac{t_x}{u_x(t_x,t_y,t_z)} \right)^s \nonumber \\ & &\left( \frac{t_y}{u_y(t_x,t_y,t_z)} \right)^{(r_1-s)} \left( \frac{t_z}{u_z(t_x,t_y,t_z)} \right)^{(r_2-s)}  (1-t_x-t_y-t_z)^{(K-r_1-r_2+s)}\, . \label{pttt}
\eea
The second moment entropy is the global maximum: $s_{\rm 2nd} = {\rm
  max}_{t_x,t_y,t_z} s_{\rm 2nd}(t_x,t_y,t_z) $.

For the regular graphs $Q(l)=\delta_{l,L}$ the expressions for both the first and second moment simplify considerably. For the first moment, the inverse of (\ref{saddle_u}) is explicit $u=[t/(1-t)]^{1/L}$ and thus 
\be
  s_{\rm ann \, reg }(t) =  \frac{L}{K} \log{\left\{ \sum_{r=0}^K \delta_{A_r,1} {K \choose r}  \left[ t^r (1-t)^{K-r} \right]^{\frac{L-1}{L}}  \right\} }\, .
\ee
For the second moment the function (\ref{saddle_tw}) is also explicitly reversible and the second moment entropy for regular graphs reads
\bea
   s_{\rm 2nd,reg}(t_x,t_y,t_z)& =& \frac{L}{K} \log \Bigg\{ \sum_{r_1,r_2=0}^K \sum_{s=\max{(0,r_1+r_2-K)}}^{\min{(r_1,r_2)}}   \frac{ K! \delta_{A_{r_1},1} \delta_{A_{r_2},1}  }{ (r_1-s)! \,  (r_2-s)! \, s! \, (K-r_1-r_2+s)!} 
\nonumber \\ & &
\left[ t_x^{s} t_y^{(r_1-s)} t_z^{(r_2-s)}  (1-\sum_w t_w)^{(K-r_1-r_2+s)}  \right]^{\frac{L-1}{L}} \Bigg\}\, .
\eea

\section{The Phase Diagram}
\label{sec:phase}

\subsection{Non-locked occupation problems}

The phase diagram of the non-locked occupation problems that we have explored
is qualitatively similar to the one of K-satisfiability and graph coloring
studied recently in detail in
\cite{MezardParisi02,MezardZecchina02,KrzakalaMontanari06,ZdeborovaKrzakala07,MontanariRicci08}.
We thus only briefly summarize the main findings in order to be able to appreciate
the difference between the locked and the non-locked problems.

As one adds constraints to a typical non-locked problem the space of solutions
undergoes several phase transitions. When the density of  constraints is very small
the replica symmetric solution is correct and most of the solutions lie in one
cluster. As the density of constraints is increased, the point-to-set
correlations, defined via the reconstruction on trees, no longer decay to zero.
This is the clustering transition, at this point the space of solutions splits
into exponentially many well separated (energetically or entropically)
clusters. But as long as an exponential number of such clusters is needed to
cover almost all the solution the observables like entropy, magnetizations,
two point correlations, etc. behave as if the replica symmetric solution was
still correct. This phase is called the {\it dynamical} 1RSB. When the
constraint density is further increased the space of solutions undergoes the
so-called condensation transition. In the condensed phase only a finite number
of clusters is needed to cover an arbitrarily large fraction of solutions.
Increasing again the density of constraints, one crosses  the
satisfiability transition where all the solutions disappear.

We remind at this point that in the non-locked occupation problems, where the
sizes of clusters fluctuates, the survey propagation equations are not
equivalent to the reconstruction on trees. More technically said the 1RSB
solutions at $m=0$ and at $m=1$ are different, for example a non-trivial
solution appears at different connectivities.

A second class of important phase transitions in the space of solutions of the
non-locked problems concerns the so-called frozen variables, which might be
responsible for the onset of algorithmic hardness \cite{ZdeborovaKrzakala07}.
A variable is frozen in a cluster if in all the solutions belonging to that
cluster it takes the same value. A cluster is frozen if a finite fraction of
variables are frozen in that cluster. A solution is frozen if it belongs to a
frozen cluster. As the number of constraints is increased the clusters tend to
freeze. We define two transition points. The first one, called the {\it
  rigidity} transition \cite{ZdeborovaKrzakala07}, is defined as the point
where almost all solutions become frozen. The second one, the {\it freezing}
transition, is defined as the point where strictly all solutions become
frozen.

In the cavity method every cluster is associated with a solution of the
BP equations. A frozen variable $i$ is described by a marginal
probability (\ref{marginal}) which is either equal to $(\chi^i_0,\chi^i_1)=(1,0)$ or
 to $(\chi^i_0,\chi^i_1)=(0,1)$. The rigidity transition is then computed as the connectivity
at which such ``frozen beliefs'' $\chi$ appear in the dominating clusters.
 If this
transition happens before the condensation transition then it is given by the
onset of a nontrivial solution to the naive reconstruction, eq.~(\ref{eq:mus}). The rigidity transition was computed for the graph coloring in
\cite{ZdeborovaKrzakala07,Semerjian07}, in the bicoloring of hyper-graphs
\cite{DallAstaRamezanpour08}, or the K-SAT in
\cite{MontanariRicci08,Semerjian07}. The freezing transition
was studied with probabilistic methods in K-SAT with large
$K$ in \cite{AchlioptasRicci06} and numerically in 3-SAT in
\cite{ArdeliusZdeborova08}. 

\subsection{Locked occupation problems}
\label{sec:locked}

{\bf Point-like clusters ---} The main property which makes the locked
problems special is that every cluster consists of a single configuration and
has thus zero internal entropy. One way to show this is realizing that in the
locked problems if $\{s_i\}$ is a satisfying configuration then
\begin{subequations}
\label{BP_fixed}
\bea
\psi^{a\to i}_{s_i} = 1\, , \quad   \psi^{a\to i}_{\neg s_i} = 0\, ,\\  
\chi^{i\to a}_{s_i} = 1\, , \quad   \chi^{i\to a}_{\neg s_i} = 0   
\eea
\end{subequations}
is a fixed point of BP eqs.~(\ref{BP1}-\ref{BP2}). The
corresponding entropy is then zero, as $Z^i=Z^{a}=1$ for all $i$, $a$. In the
derivation of \cite{MezardMontanari07} the fixed points of the belief
propagation equations correspond to clusters. Thus in the locked problems
every solution may be thought of as a cluster. Such a situation was previously
encountered in a few problems  \cite{KrauthMezard89b,MartinMezard04,MartinMezard05} and called
the {\it frozen} 1RSB because all
the variables, clusters and solutions are frozen in such a case.

{\bf The clustering transition ---} 
In terms of the reconstruction on trees the
situation in the locked problems is trivial because the boundary conditions on
leaves always imply uniquely all the variables in the body of the tree and
also the root. However one may ask what happens if the assignment of a small fraction of
the variable on leaves is also forgotten -- we call this the {\it small noise
  reconstruction} on trees \footnote{The related, but different, concept of {\it robust reconstruction} was studied in \cite{JansonMossel04}. Whereas in the small noise reconstruction one investigates the effect of an infinitesimal noise-rate in the robust reconstruction one studies the effect of a noise-rate arbitrarily close to 1.}.
 In the non-locked problems nothing changes. In the
locked problems the small noise reconstruction is not equivalent to the
reconstruction. At sufficiently small connectivities the small noise
reconstruction is not possible, that is if we introduce a small noise in the
leaves all the information about the root is lost. In the same spirit: we
showed that every solution corresponds to a fixed point of the belief
propagation of the type (\ref{BP_fixed}), but we did not ask if such a fixed point is
stable under small perturbations. If an infinitesimal fraction of messages in
(\ref{BP_fixed}) is changed, will the iterations (\ref{BP1}-\ref{BP2}) 
converge back to the unperturbed fixed point or not? Again for sufficiently small
connectivity it will not. This leads us back to a definition of the clustering
transition which needs to be refined for the locked problems.

We thus define the {\it clustering transition} as the threshold for the small
noise reconstruction. As all the clusters are frozen the reconstruction
problem is equivalent to the naive reconstruction which deals only with the
frozen variables. So for example on the ensemble of random regular graphs it
is sufficient to investigate the stability of the solution
$\mu_0=\mu_1=1$ of eqs.~(\ref{eq:mu1}-\ref{eq:mu0}) under iteration. It is immediate to see
that if $L\ge 3$ then the solution $\mu_1=\mu_0=1$ of
(\ref{eq:mu1}-\ref{eq:mu0}) is always iteratively stable. When $L=2$ we
observed empirically that the solution $\mu_1=\mu_0=1$ is not stable and the
only other solutions is $\mu_1=\mu_0=0$. Thus in the regular graphs ensemble
of the locked problems the clustering transition is at $L=3$.

For a general graph ensemble it is simpler to realize that as the internal entropy of clusters is zero the 1RSB solution does not depend on the value of the Parisi parameter $m$. Thus in particular the small noise reconstruction is equivalent to the iterative stability of the BP-like fixed point of the survey propagation equations.  

We have found that, in the locked occupation problems, the SP equations (\ref{eq:SP1}-\ref{eq:SP2}), when initialized randomly, have two possible iterative fixed points:
\begin{itemize}
\item The trivial one: $q^{a\to i}_*=p^{i\to a}_*=1$, $q^{a\to i}_1=p^{i\to a}_1=q^{a\to i}_{0}=p^{i\to a}_{0}=0$ for all edges $(ai)$.
\item The BP-like one: $q^{a\to i}_*=p^{i\to a}_*=0$, $q^{a\to i}=\psi^{a\to i}$, $p^{i\to a}=\chi^{i\to a}$ for all edges $(ai)$, where $\psi$ and $\chi$ is the solution of the BP equations (\ref{BP1}-\ref{BP2}) found with high probability by iterating the equations from a uniformly random initial condition.
\end{itemize}
The small noise reconstruction is then investigated, using the population dynamics, from the  stability of the BP-like fixed point under iteration. If it is stable then the small noise reconstruction is possible and the phase is clustered. If it is not stable then we are in the liquid phase.
From a geometric point of view, we conjecture that in the liquid phase the Hamming distance separation between solutions grows only proportionally to $\log N$; on the contrary, when small noise reconstruction is possible we expect this Hamming distance to be extensive (proportional to $N$).

{\bf The satisfiability threshold ---} The BP
eqs.~(\ref{BP1}-\ref{BP2}) have many fixed points. However, when we solve them
iteratively starting from a random initial condition we always find the same
fixed point which does not correspond to a satisfying assignment
(\ref{BP_fixed}). We call this fixed point and its corresponding entropy the
replica symmetric solution. It should actually be thought of as a
fixed point of the survey propagation equations as explained in the previous
paragraph. The important fact is that it gives the correct
entropy (\ref{eq:entropy_l}), and also the correct marginal probabilities.

The satisfiable threshold in the locked problems is then computed as the
average connectivity $l_s$ at which the replica symmetric entropy
(\ref{eq:entropy_l}) decreases to zero \cite{KrauthMezard89b}, $s(l_s)=0$. This is the first of many
quantities in the locked problems which can be computed with much smaller
effort then in the non-locked problems. The condensed phase, where the space
of solutions is dominated by a finite number of clusters does not exist in the
locked problems, and the condensation transition coincides with the
satisfiability threshold. 

\begin{figure}[!ht]
  \resizebox{0.6\linewidth}{!}{
  \includegraphics{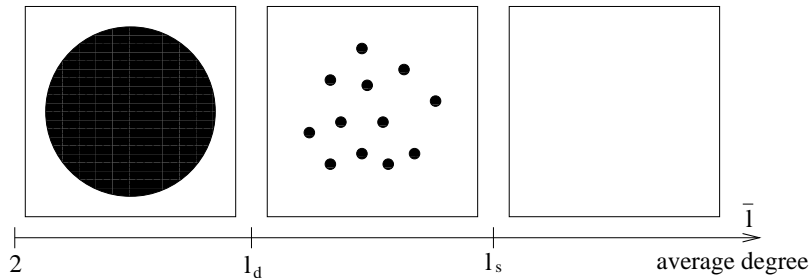}}
  \caption{\label{Fig:phase_diag} Sketch of the phase diagram in the locked problems. At low constraint density $\overline l<l_d$ the solutions are separated by logarithmical distance but if any sort of noise is introduced this separation disappears. In the clustered phase $l_d \le \overline l <l_s$ the space of solutions is made of well separated single solutions. And eventually the satisfiability transition $l_s$ comes beyond which solutions do not exist.}
\end{figure}
{\bf Summary of the phase diagram ---} In contrast to the zoo of phase transitions in non-locked problems, in the
locked problems we find only three phases, sketched in
Fig.~\ref{Fig:phase_diag}, the critical connectivity values are given in
Table~\ref{tab:locked}.
\begin{itemize}
   \item{The liquid phase, for connectivities $\overline l<l_d$: In this phase the small noise reconstruction is not possible. Equivalently the BP-like iterative fixed point of the survey propagation equations is not stable. If one considers the problem at a very small temperature, the 1RSB equations have only the trivial solution, such a situation was observed previously in the perfect matching problem \cite{ZdeborovaMezard06,ZhouYang03}.
 We expect that the Hamming distance separation between solutions in this phase is only logarithmic.}
   \item{The clustered phase, for $l_d<\overline l<l_s$: In this phase the small noise reconstruction is possible. The BP-like iterative fixed point of the survey propagation equations is stable. The 1RSB equations have a non-trivial solution even at an infinitesimal temperature. We expect that the solutions are separated by an extensive Hamming distance, in other words there is a gap in the weight enumerator function, just like in the XOR-SAT \cite{MoraMezard06}. This property is crucial in low density parity check codes \cite{DiRichardson06}.}
   \item{The unsatisfiable phase, for $\overline l>l_s$: no more solutions exist.}
\end{itemize}
All the other phase transitions we described for the non-locked problems have become very simple: The clustering transition coincides with the rigidity and freezing. And the satisfiability transition coincides with the condensation one. 

\begin{table}[!ht]
\begin{center}
\begin{tabular}{|l|l||l||l|l|l|l|} \hline
A       & name      &\, $L_{s}$\, &\, $c_{d}$&\, $c_{s}$&\, $ l_d$&\, $ l_s$\\ \hline 
0100    & 1-in-3 SAT       & 3   &0.685(3)\, &0.946(4)\, &2.256(3)\, &2.368(4)\,  \\ \hline 
01000   & 1-in-4 SAT       & 3   & 1.108(3)  & 1.541(4)  & 2.442(3)  & 2.657(4)   \\ \hline 
00100*  & 2-in-4 SAT       & 3   & 1.256     & 1.853     & 2.513     & 2.827      \\ \hline 
01010*  & 4-odd-PC        & 5   & 1.904     & 3.594     & 2.856     & 4          \\ \hline 
010000  & 1-in-5 SAT       & 3   & 1.419(3)  & 1.982(6)  & 2.594(3)  & 2.901(6)   \\ \hline 
001000  & 2-in-5 SAT       & 4   & 1.604(3)  & 2.439(6)  & 2.690(3)  & 3.180(6)   \\ \hline 
010100  & 1-or-3-in-5 SAT  & 5   & 2.261(3)  & 4.482(6)  & 3.068(3)  & 4.724(6)   \\ \hline 
010010  & 1-or-4-in-5 SAT  & 4   & 1.035(3)  & 2.399(6)  & 2.408(3)  & 3.155(6)   \\ \hline 
0100000 & 1-in-6 SAT       & 3   & 1.666(3)  & 2.332(4)  & 2.723(3)  & 3.113(4)   \\ \hline 
0101000 & 1-or-3-in-6 SAT  & 6   & 2.519(3)  & 5.123(6)  & 3.232(3)  & 5.285(6)   \\ \hline
0100100 & 1-or-4-in-6 SAT  & 4   & 1.646(3)  & 3.366(6)  & 2.712(3)  & 3.827(6)   \\ \hline
0100010 & 1-or-5-in-6 SAT  & 4   & 1.594(3)  & 2.404(6)  & 2.685(3)  & 3.158(6)   \\ \hline
0010000 & 2-in-6 SAT       & 4   & 1.868(3)  & 2.885(4)  & 2.835(3)  & 3.479(4)   \\ \hline 
0010100*& 2-or-4-in-6 SAT  & 6   & 2.561     & 5.349     & 3.260     & 5.489      \\ \hline
0001000*& 3-in-6 SAT       & 4   & 1.904     & 3.023     & 2.856     & 3.576      \\ \hline 
0101010*& 6-odd-PC        & 7   & 2.660     & 5.903     & 3.325     & 6          \\ \hline
\end{tabular}  
\caption{\label{tab:locked}The locked cases of the occupation CSPs for $K\le
  6$. In the regular graphs ensemble the space of solutions is clustered for $L \ge L_d=3$,
  and the problem is unsatisfiable for $L\ge L_s$. The values $c_d$ and $c_s$ are the
  critical parameters of the truncated Poissonian ensemble (\ref{Poiss}), the
  corresponding average connectivities $l_d$ and $l_s$ are
  given via eq.~(\ref{l_aver}). In all there problems the replica symmetric solution is stable at least up
  to the satisfiability threshold. The balanced cases are marked as *, their
   dynamical threshold follows from (\ref{cd_bal}), and their satisfiability
  threshold can be computed from the second moment method.}
\end{center}
\end{table}

Finally we would like to mention the stability of the frozen 1RSB solution
towards more levels of replica symmetry breaking. In more geometrical terms,
one should check whether the solutions do not tend to aggregate into clusters.
This is called stability of type I in the literature
\cite{MontanariRicci03,RivoireBiroli04,KrzakalaZdeborova07}. In the locked
problems it is equivalent to the finiteness of the spin glass susceptibility.
In all the locked occupation problems we have studied, including all those
in Tab.~\ref{tab:locked}, we have seen that the frozen 1RSB solution
is always stable in the satisfiable phase and sometimes becomes unstable at a point in
the unsatisfiable phase. This means that our description of the satisfiable
phase, and the determination of the thresholds $l_d$ and $l_s$, should be exact.

\subsection{The balanced LOPs}
\label{sec:bal}

We have seen that the phase diagram in the locked
problems is much simpler than in the more studied constraint satisfaction
problems as the K-SAT or coloring. In this section we describe a subclass of
the locked problems --- the so-called {\it balanced} locked problems --- where
the situation is even simpler. In particular, the clustering and the satisfiability threshold can be determined easily, and the second moment method can be used to prove rigorously the validity of this determination of the satisfiability threshold. This makes the balanced locked problems very
interesting from the mathematical point of view.

The {\it balanced} occupation problems are defined via the property that 
two random solutions are almost surely at Hamming distance $N/2+o(N)$.
This property may of course depend on the connectivity distribution
$Q(l)$. A necessary condition for the problem to be balanced it that the
vector $A$ be palindromic, meaning that $A_{r}=A_{K-r}$. But not all the palindromic
problems are balanced, the simplest such example is the 1-or-4-in-5 SAT,
$A=010010$, where the symmetry is spontaneously broken in the same way as in a
ferromagnetic Ising model.

As we argued in the previous section in the locked problems the replica
symmetric approach (BP) gives the exact marginal probabilities
and total entropy. Therefore a problem is balanced if and only if the
iterative fixed point of the BP equations
(\ref{BP1}-\ref{BP2}) is such that all the beliefs are equal to $1/2$.

We do not know of any simpler general rule to decide if a problem is balanced. For
$K\le 12$, there is no exception to the following empirical rule:  all the
problems which can be obtained from a Fibonacci-like recursion
\be
      0A_{K}0 = A_{K+2}  \quad \quad 01A_K10 = A_{K+4}
\ee 
from $A_2=010$ or $A_4=01010$ are balanced in their satisfiable phase. There are, however, other balanced locked problems which cannot be obtained this way, the simplest example is $A=0001001000$.

{\bf Clustering threshold in the balanced LOPs}:
 The clustering threshold is
given by the small noise reconstruction, i.e. by the stability of the naive
reconstruction procedure as explained in Sec.~\ref{sec:locked}. In balanced LOPs,
the messages are symmetric, $\psi_0=\psi_1=1/2$, and thus also the probability for the root variable to be uniquely determined by the leaves is independent of the value which has been broadcast:
$\mu_0=\mu_1=\mu$. For a graph ensemble with excess degree distribution
$q(l)$, one can write explicitly the self-consistency condition on $\mu$:
\be 
\mu = \frac{2}{g_A} \sum_{r=0}^k  \delta_{A_{r+1},1} \delta_{A_r,0} {k\choose r} \sum_{s=0}^{s_1} {r \choose s} \left[
  1- \sum_{l=0}^\infty q(l) (1-\mu)^l \right]^{k-s} \left[ \sum_{l=0}^\infty
  q(l) (1-\mu)^l \right]^{s} \, , \label{eq:mu} 
\ee 
where $k=K-1$, and $g_A = \sum_{r=0}^k \delta_{A_{r+1},1} {k \choose r}+ \sum_{r=0}^k \delta_{A_r,1} {k
  \choose r}$. For the ensemble of graphs with truncated Poissonian degree
distribution of coefficient $c$ we derive from (\ref{excess})
 \be 
\mu = \frac{2}{g_A}
\sum_{r=0}^k \delta_{ A_{r+1},1}\delta_{ A_r,0 } {k\choose r} \sum_{s=0}^{s_1} {r
  \choose s} \left( \frac{1 - e^{-c\mu}}{1-e^{-c}}\right)^{k-s}
\left(\frac{e^{-c\mu}-e^{-c}}{1-e^{-c}}\right)^{s} \, . 
\ee
 The clustering threshold is defined as the value of $c$ where the fixed point $\mu=1$ becomes unstable. One gets:
\be
 \frac{e^{c_d}-1}{c_d}=K-1 -\frac{\sum_{r=0}^{K-2} 
\delta_{A_{r+1,1}}\, \delta_{A_{r-1},0}\, \delta_{A_{r},0}\,  {K-1 \choose r}}
{\sum_{r=0}^{K-2} \delta_{A_{r+1},1} \, {K-1 \choose r}}\, .
\label{cd_bal}
\ee
These values are summarized in Table \ref{tab:locked}, where the balanced locked problems are marked by a $\ast$.

{\bf Satisfiability threshold in the balanced LOPs:}
For the balanced locked problems the replica symmetric entropy is given by:
\be
s_{\rm sym}(\overline l)=\log{2}+\frac{\overline l}{K}
\log{\left[2^{-K}\sum_{r=0}^K \delta_{A_r,1} {K\choose r}\right] }\label{eq:s_sym}
\, ,        
\ee
where $\overline l$ is the average degree of variables.
Notice the simple form of this entropy: in the balanced locked problems each added constraint destroys a fraction of solutions exactly equal to the fraction of configurations that satisfy a single constraint.
The satisfiability threshold is then given by the point $l_s$ where this entropy is zero. 

{\bf Second moment method in the balanced LOPs:}
In all the balanced LOPs that we have considered
we found numerically that  the second moment entropy,
(\ref{2nd_d}), is exactly twice the annealed entropy (\ref{1st_d}), $2s_{\rm
  ann}=s_{\rm 2nd}$. A hint that this may happen comes from the following observations: 
\begin{itemize}
  \item{The annealed entropy (\ref{1st_d}) has a stationary point at $t=1/2$ ($u=1$, $x=1$). At this stationary the entropy evaluates to (\ref{eq:s_sym}).}
  \item{The second moment entropy (\ref{2nd_d}) has a stationary point at
      $t_x=t_y=t_z=1/4$ ($u_x=u_y=u_z=1$, $x=y=z=1$). At this stationary point
      the second moment entropy evaluates to twice the annealed entropy
      (\ref{eq:s_sym}). This can be seen using the Vandermonde's combinatorial
      identity 
\be {K \choose r_2} = \sum_{s=0}^{r_1} {r_1 \choose s} {K-r_1
        \choose r_2 -s}\, . 
\ee}
\end{itemize}
We checked numerically that in the balanced LOPs the global maxima of $s_{\rm ann}$ and $s_{\rm 2nd}$ is always given by these stationary points (the second moment entropy has another stationary point at $t_x=1/2, t_y=t_z=0$ or $t_x=0, t_y=t_z=1/2$, but at this point it is equal to the first moment entropy at $t=1/2$). On the contrary, in the non-locked or non-balanced problems we always found another competing maximum. 

If one accepted the result $2s_{\rm ann}=s_{\rm 2nd}$, and made the reasonable
assumption that the satisfiability threshold is sharp, then Chebyshev's
inequality (\ref{Cheb}) would prove the correctness of the satisfiability
threshold computed from (\ref{eq:s_sym}). 
Therefore the full class of balanced LOPs is a candidate for a rigorous mathematical determination of
the satisfiability threshold. This would be quite interesting,
 as it would noticeably enlarge 
the list of problems where the threshold is known rigorously (so far 
  only a handful of 
sparse NP-complete CSPs are in this category : the 1-in-$K$ SAT
\cite{AchlioptasChtcherba01,RaymondSportiello07}, the $2+p$-SAT
\cite{CoccoDubois03,AchlioptasKirousis01} and the $(3,4)$-UE-CSP
\cite{ConnamacherMolloy04}).

Let us summarize qualitatively what are the main features of the balanced locked
occupation problems that make the fluctuations of the number of solutions so small that the second
moment method presumably gives the exact satisfiability threshold.
\begin{itemize} 
    \item{Balancing --- It is well known that the second moment method works better if most of the weight is on the most numerous configurations (that is the half-filling ones). In the K-SAT problem several reweighting schemes were introduced in order to improve the second moment lower bounds \cite{AchlioptasMoore06,AchlioptasNaor05}. This is also the reason why the second moment bound is much sharper in the balanced NAE-SAT (bicoloring) than in the K-SAT \cite{AchlioptasMoore06}.}
    \item{Reducing fluctuations in the connectivity --- Naturally, reducing the fluctuations of the variables connectivity reduces the fluctuation of the number of solutions. Our work shows that the necessary step is not to have leaves. Fluctuating higher degrees do not really pose a problem.}
    \item{Locked nature of the problem --- And finally the key point is the locked structure of the problem. It was remarked in \cite{ArdeliusZdeborova08} that the clusters-related quantities fluctuate much less than the solutions-related ones. Thus the fact that clusters do not have a fluctuating size, but size $1$ is the crucial property needed to make the second moment method sharp. This is exactly what happens in the XOR-SAT problem, where the second moment becomes exact when it is restricted to the 'core' of the graph \cite{MezardRicci03,CoccoDubois03}}
\end{itemize}

\section{Numerical studies}
\label{sec:alg}

We shall show in this section that the LOPs, in their
whole clustered phase, seem to be very hard from the algorithmic point of view. We shall illustrate this by testing and
analyzing the performance of some of the best algorithms developed for
random 3-satisfiability, the canonical hard constraint satisfaction problem.

Our first study uses a complete algorithm and shows that, like in other
problems such as satisfiability and coloring, the hardest instances are found in the neighborhood of the satisfiability transition. We then turn to incomplete algorithms, which are aimed at finding a SAT configuration 
when it exists.

The best performance for incomplete algorithms is nowadays attributed to the survey propagation inspired
decimation \cite{MezardZecchina02,Parisi03} and the survey propagation
inspired reinforcement \cite{ChavasFurtlehner05}. In the random 3-SAT problem both of these algorithms were
reported to work in linear time (or at most log-linear time) up to a
constraint density about $\alpha=4.252$, to be compared with the
satisfiability threshold $\alpha_s=4.267$ and the clustering transition
$\alpha_d=3.86$.

As we saw in Sec.~\ref{sec:locked}, in the LOPs, the survey propagation algorithm has no
advantage over the belief propagation algorithm. We thus study the performance of the BP
inspired decimation and reinforcement. The conclusions are as follows: the BP decimation fails
in the LOPs even at very low connectivities; the BP reinforcement
works in linear time in the non-clustered phase but fails in the clustered phase. 

\subsection{Exhaustive search results}

One way to solve a LOP is to transform it into a conjunctive normal form
(CNF), and use some of the open source complete solvers of the satisfiability
problem. We have done such a study for the 1-or-3-in-5 SAT problem. We have
generated random instances of this problem from a truncated Poisson ensemble,
with $M$ constraints. Each instance has been transformed into a satisfiability
formula by mapping every constraint into $\sum_{r=0}^K \delta_{A_r,0} {K
  \choose r}$ CNF clauses: for every constraint of $K$ variables, one creates
as many CNF clauses, out of the $2^K$ possible clauses, as there are forbidden
configurations. We have applied a branch-and-bound based open-source SAT solver
called {\tt MiniSat 1.14} \cite{EenSorensson04} to test the satisfiability and
to compute the running time needed by this algorithm to decide the satisfiability.

\begin{figure}[!ht]
  \resizebox{\linewidth}{!}{
  \includegraphics{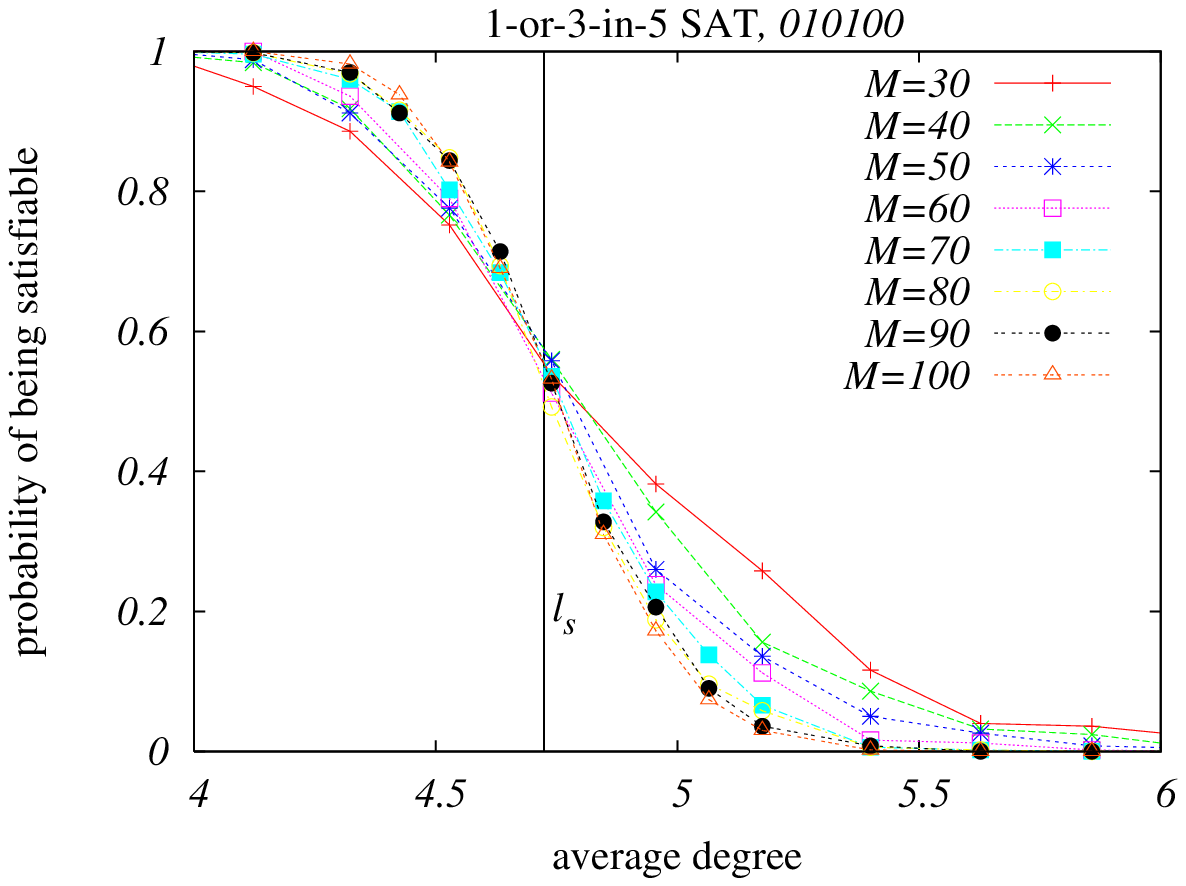}
  \includegraphics{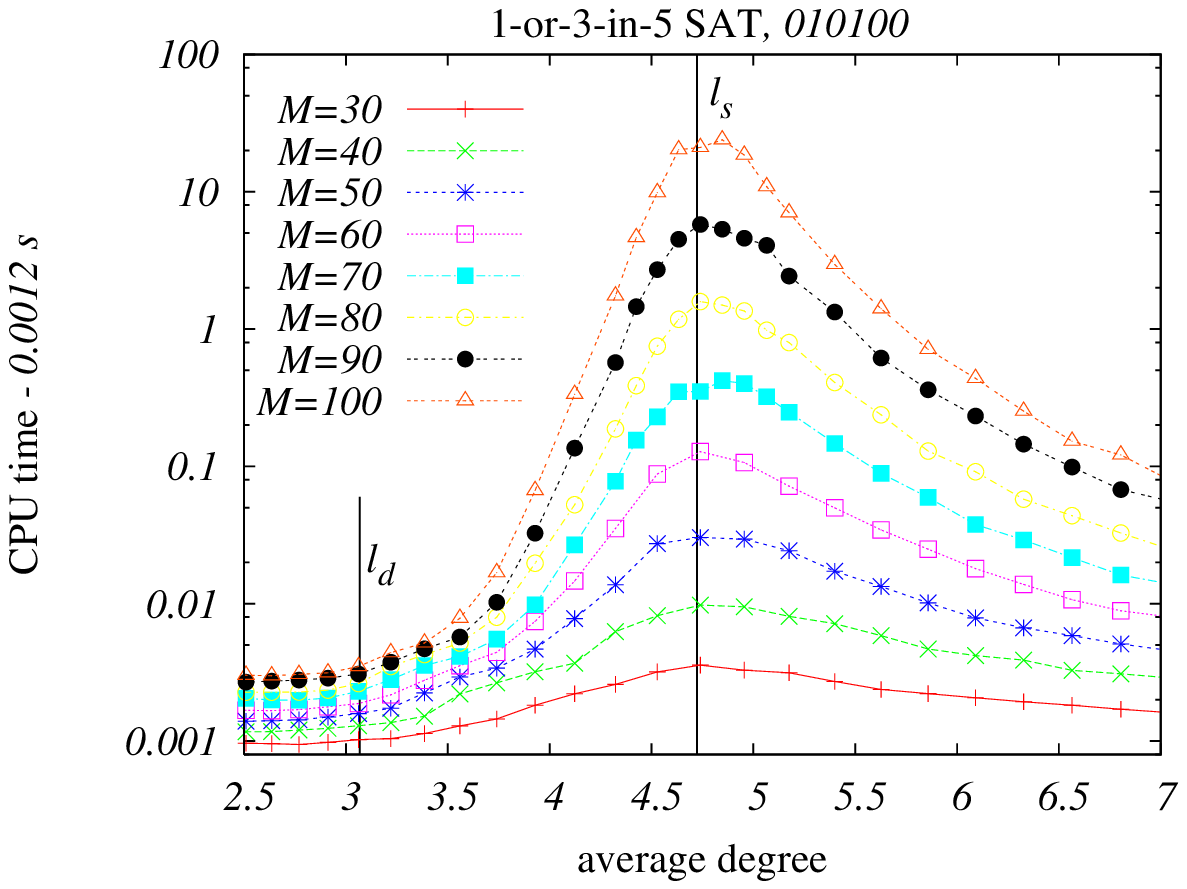}}
  \caption{\label{Fig:complete_algo} \emph{Left}: Probability that a random instance of the 1-or-2-in-5 SAT is satisfiable, versus the average degree. The probability is computed from  $500$ instances  generated from the truncated Poisson ensemble. The vertical line shows the analytical prediction of the
 value of the satisfiability transition.
\emph{ Right}: Median over the same $500$ instances of the CPU running time of the complete algorithm 
{\tt MiniSat 1.14}
(we have subtracted  $0.0012$ seconds from the CPU time, as this is approximately where is extrapolates for small average degree and zero system size). Alternatively one could plot the number of backtracking steps, which has a qualitatively identical behavior. 
}
\end{figure}

Figure \ref{Fig:complete_algo}, left, shows the probability that an instance
is satisfiable, plotted versus the average degree. It displays the typical
behavior of a phase transition rounded by finite size effects. Figure
\ref{Fig:complete_algo}, right, shows the median value of the CPU time  which was used
to solve an instance of the decision problem on a 2GHz MacBook laptop (note the logarithmic
scale), plotted versus the average degree. The hardest instances appear around
the satisfiability threshold $l_s$, and the time needed by the algorithm in
this region clearly grows exponentially with the size. Hard satisfiable
instances start to appear around $\overline l\simeq l_d$, although it is
difficult to assert from this data where the exponential behavior really
starts. For larger system sizes it seems that the exponential behaviour starts
way below the dynamical threshold  $l_d$.

The data shows the same qualitative behavior as has been found in similar
studies of satisfiability, with the difference that the relative width,
$(l_s-l_d)/l_s$, of the clustered phase is larger in this case than it is in
the $K=3$ or $K=4$ satisfiability problems. The existence of LOPs with such a
broad clustered phase is an appealing feature for numerical studies. In the
following sections of this paper we argue that in the locked problems the easy-hard
algorithmic threshold for the best-known incomplete solvers coincides with the
clustering transition $l_d$.

\subsection{Decimation fails in LOPs}

In BP inspired decimation one uses the knowledge of the marginal probabilities
estimated from  BP in order to identify the most biased variable, fix it to
its most probable value, and reduce the problem. Such an algorithm usually
works well even in the clustered region (for performance in K-SAT and coloring
see \cite{KrzakalaMontanari06,ZdeborovaKrzakala07}). In the locked occupation
problems the BP decimation fails badly. For example in the 1-or-3-in-5 SAT
problem, on the truncated Poisson graphs with $M=2\cdot10^4$ constraints, the
probability of success is about 25\% at $\overline l =2$, and less than 5\% at
$\overline l =2.3$, way below the clustering threshold $ l_d\simeq 3.07$.
Interestingly, the precursors of the failure of the BP decimation algorithm
observed for instance in graph coloring are not present in the locked
problems. In particular the BP equations converge during all the process and
the normalizations in the BP equations (\ref{BP1}-\ref{BP2}) stay finite.

Although we do not know how to analyze directly the BP decimation process, the
mechanisms explaining the failure of the decimation strategy can be understood
using the approach of \cite{MontanariRicci07}. The idea is to analyze an
idealized decimation process, where the variable to be fixed is chosen
uniformly at random and its value is chosen according to its exact marginal
probability. If its value is chosen according to the BP marginal we speak about the uniform BP decimation. If BP would give a fair approximation to the exact marginal throughout the decimation process, the uniform BP decimation should be equivalent to the ideal decimation. In the ideal decimation, the reduced problem obtained after $\theta N$ steps is statistically equivalent to the reduced problem created by choosing a solution uniformly at random and revealing a fraction~$\theta$ of its variables, which we now analyze, following the lines of \cite{MontanariRicci07}.

Given an instance of the CSP, consider a solution taken uniformly at random
and reveal the value of each variable with probability $\theta$. Denote $\Phi$
the fraction of variables which either  have been revealed or are directly implied
by the revealed ones. We can compute $\Phi(\theta)$ using the replica symmetric
cavity method (which is correct in the satisfiable phase of locked problems) as
follows.

Denote by $\phi^{i\to b}_s$ the probability that a variable $i$ is implied
conditioned on the value $s$ of the variable $i$ and on the absence of the
edge $(ib)$; denote by $q_s^{a \to i}$ the probability that constraint $a$
implies variable $i$ to be $s$ conditioned on:  (1) variable $i$ takes the
value $s$ in the solution we chose, (2) variable $i$ was not revealed directly
and
(3) the edge $(ai)$ is absent.
Then $\phi^{i\to b}_s$ is given by:
 \be 
\phi^{i\to b}_s = \theta + (1-\theta) \left[
  1 - \prod_{a\in \partial i - b}(1-q_s^{a \to i} ) \right]\, , 
\label{eq:qtophi}
\ee
meaning that the variable $i$ was either revealed or not, and if not it is
implied if at least one of the incoming constraints implies it. We shall write
the expression for $q_s^{a \to i}$ only for occupation problems on random
regular graphs where the replica symmetric equation is factorized. Then
$q_s^{a \to i}$ and $\phi^{i\to b}_s$ are independent of $a,b,i$: $q_s^{a \to
  i}=q_s$ and $\phi^{i\to b}_s=\phi_s $. The conditional probability $q_s$ is
the ratio of the probability that variable $i$ takes the value $s$ and is
implied by the constraint $a$ on one hand, and the probability that variable
$i$ takes the value $s$ on the other hand:
\begin{subequations}
\label{eq:qs}
\bea
   q_1 &=& \frac{1}{\psi_1 Z^{\rm reg}}\sum_{r=0}^k \delta_{A_r,0}\delta_{A_{r+1},1} {k \choose r} (\psi_1)^{lr}  (\psi_0)^{l(k-r)} \sum_{s=0}^{s_1} {r \choose s} \phi_0^{k-r} \phi_1^{r-s} (1-\phi_1)^{s} \, ,  \label{eq:q1}\\
   q_0 &=& \frac{1}{\psi_0 Z^{\rm reg}}\sum_{r=0}^k \delta_{A_r,1} \delta_{A_{r+1},0}
 {k \choose r} (\psi_1)^{lr}  (\psi_0)^{l(k-r)}\sum_{s=0}^{s_0} {k-r \choose s} \phi_1^{r} \phi_0^{k-r-s} (1-\phi_0)^{s}\, ,\label{eq:q0} 
\eea
\end{subequations}
where $l=L-1$, $k=K-1$. The sum over $r$ goes over all the possible numbers of
$1$'s being assigned on the incoming variables, and the numbers
$\psi_0,\psi_1$ are the cavity probabilities, solutions of the BP equations
(\ref{RS1}-\ref{RS2}). The indices $s_1,s_0$ in the second sum of both
equations are the largest possible but such that $s_1\le r$, $s_0\le K-1-r$,
and $\sum_{s=0}^{s_1}A_{r-s}=0$, $\sum_{s=0}^{s_0}A_{r+1+s}=0$. The terms $\phi^{k-r}_0 \phi^{r-s}_1
(1-\phi_1)^s$ and
$\phi^r_1 \phi^{k-r-s}_0 (1-\phi_0)^s$  are the probabilities that a sufficient number of
incoming variables was revealed such that the out-coming variable is implied
(not conditioned on its value). $Z^{\rm reg}$~is the
normalization in (\ref{RS1}-\ref{RS2}).

Iterations of eqs.~(\ref{eq:qtophi}-\ref{eq:qs}) with the  initial condition $\phi=\theta$ give us the fixed point for $q_0,q_1$. 
The total probability that a variable is fixed is then computed as
\be 
\Phi(\theta) = \theta + (1-\theta)
\left\{ \mu_1 [1- (1- q_1)^L ] + \mu_0 [1- (1- q_0)^L ] \right\}\,
, \label{eq:phi_th} 
\ee
where $\mu_0,\mu_1$ are the total BP marginals,
$\mu_s=\psi_s^L/(\psi^L_0+\psi^L_1)$. Notice the analogy between
eqs.~(\ref{eq:q0}-\ref{eq:q1}) and the equations for the naive reconstruction
(\ref{eq:mu0}-\ref{eq:mu1}).

\begin{figure}[!ht]
  \resizebox{\linewidth}{!}{
  \includegraphics{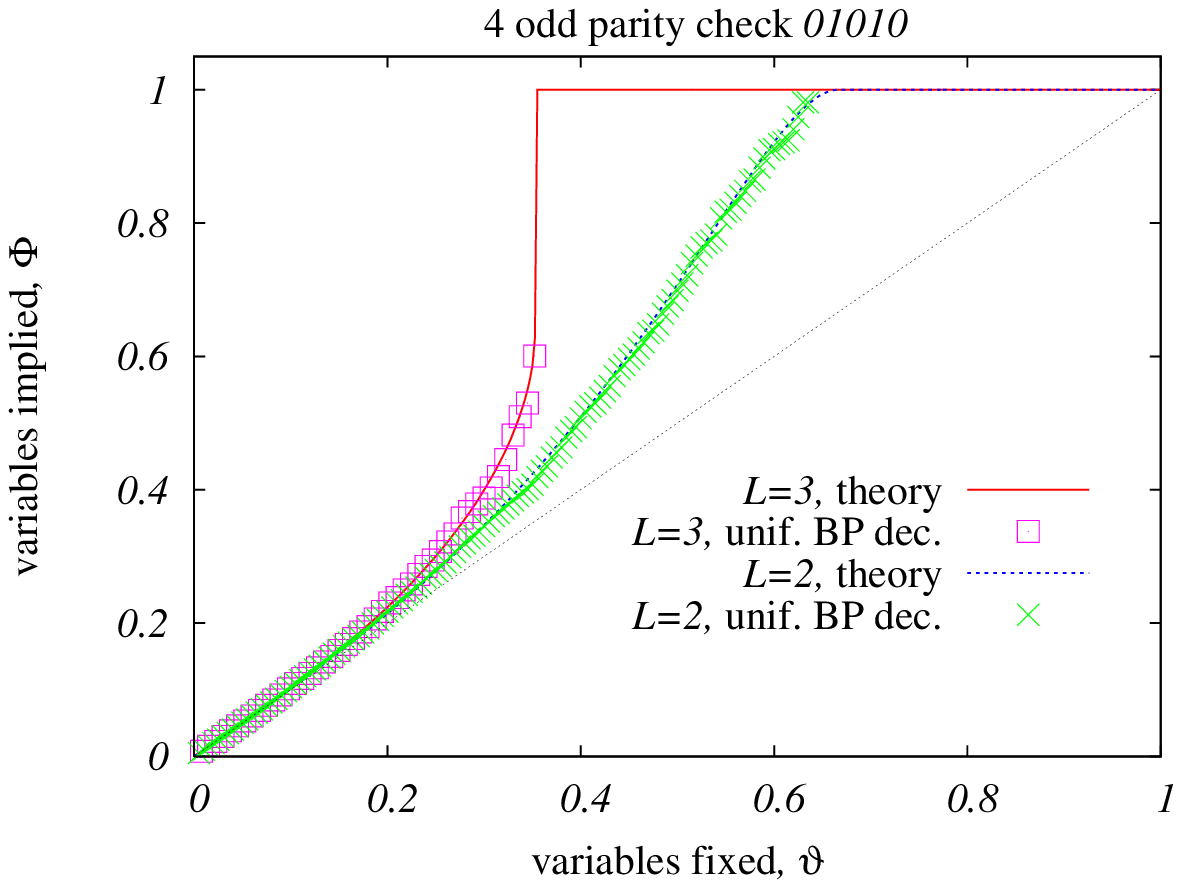}
  \includegraphics{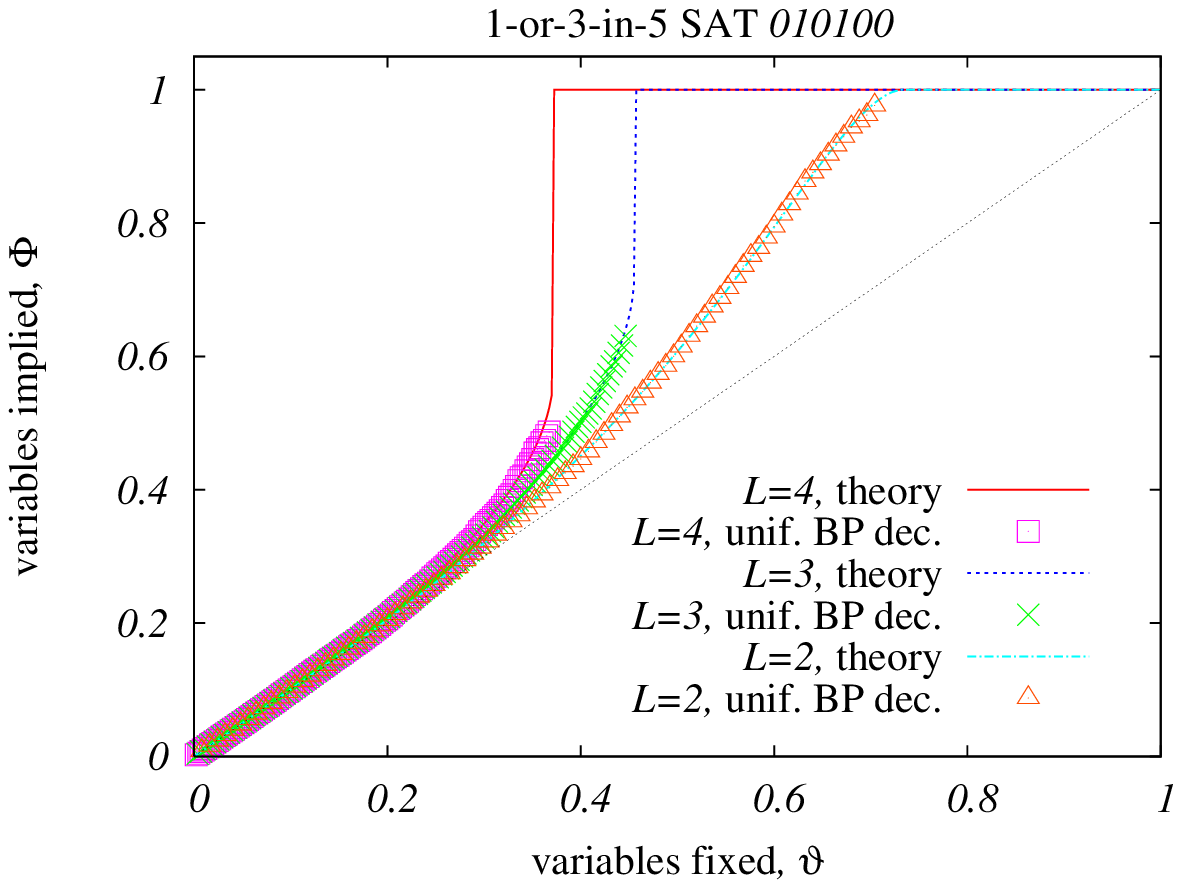}}
  \caption{\label{Fig:BP_anal} Analytical and numerical study of the BP inspired uniform decimation. The number of variables which are directly implied, $\Phi(\theta)$, is plotted against the number of fixed variables $\theta$ in two of the LOPs on the regular graph ensemble with connectivity $L$.
}
\end{figure}

\begin{figure}[!ht]
  \resizebox{\linewidth}{!}{
  \includegraphics{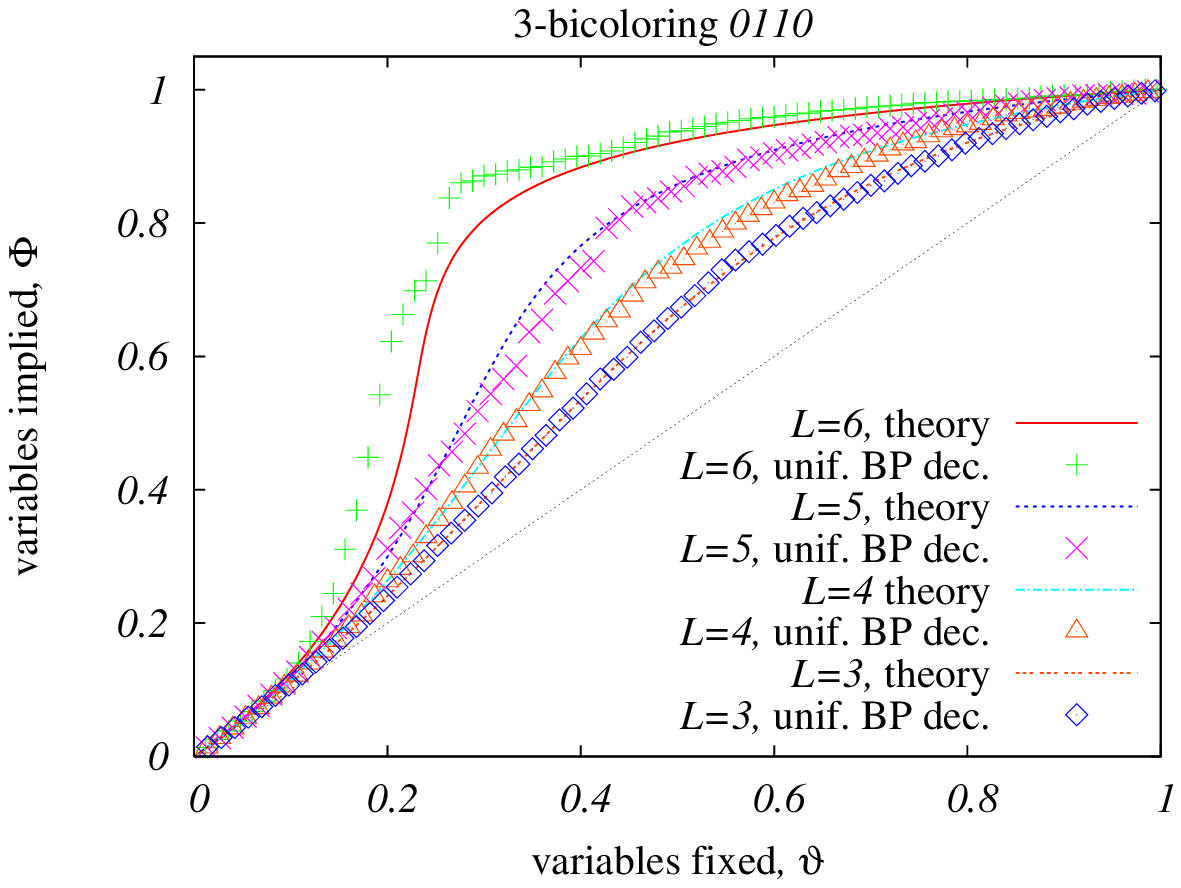}
  \includegraphics{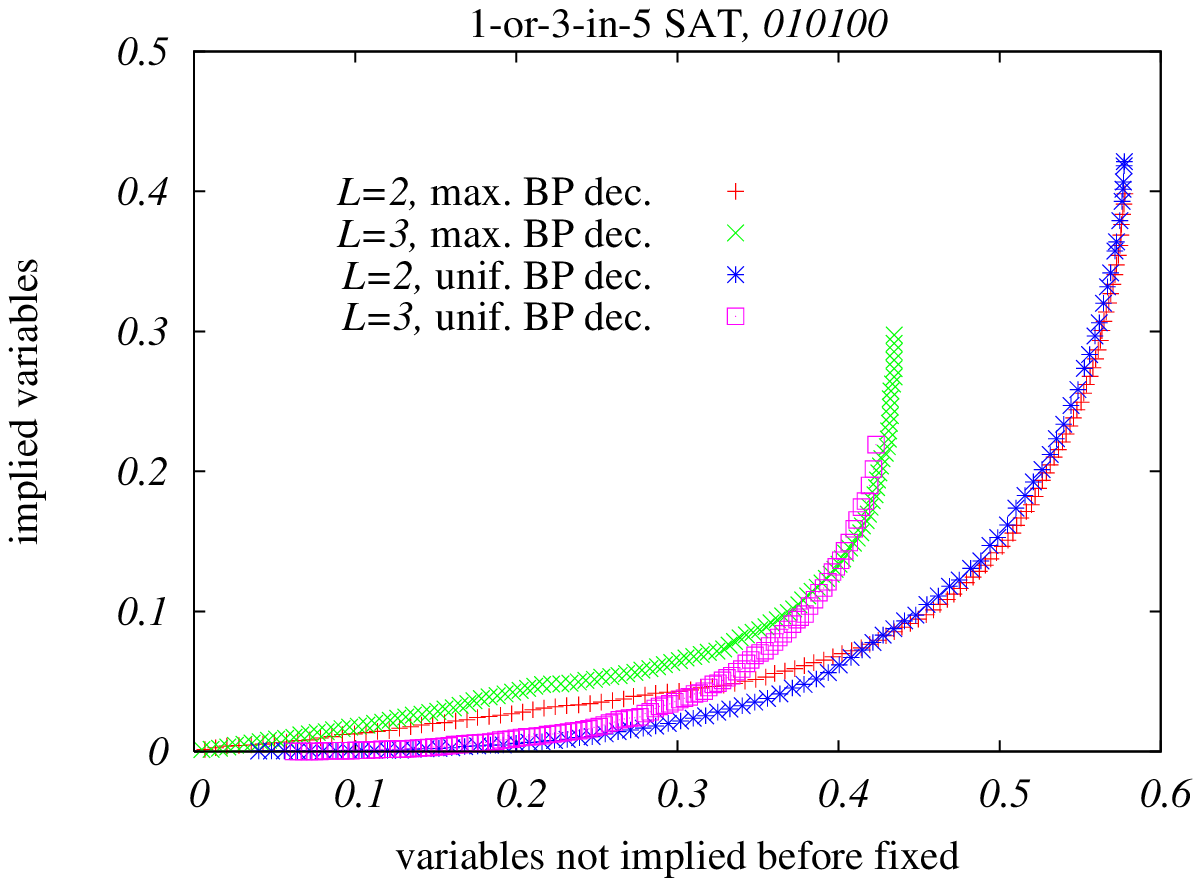}}
  \caption{\label{Fig:BP_dec} Left: For comparison, the BP uniform decimation works well on the non-locked problems, the example is for bicoloring. Right: Comparison of the uniform BP decimation with the maximal BP decimation. The number of variables which are directly implied or were directly implied before being fixed is plotted against the number of variables which were free just before being fixed. The behavior of the two decimation strategies is similar. The divergence of the derivative of this function marks the point of failure.}
\end{figure}

In Fig.~\ref{Fig:BP_anal} we compare the function $\Phi(\theta)$ obtained from
the analytical study of ideal decimation (\ref{eq:phi_th}) with the
experimental performance of the uniform BP decimation. Before the failure of
the decimation algorithm (when a contradiction is encountered) the two curves
are in very good agreement. This study shows two different reasons for the
failure of the algorithm:
\begin{itemize}
\item{Avalanche of direct implications -- In some cases the function
    $\Phi(\theta)$ has a discontinuity at a certain spinodal point $\theta_s$
    (e.g. $\theta_s\approx 0.46$ at $L=3$ for the 1-or-3-in-5 SAT problem). For
    $\theta<\theta_s$,  fixing one variable generates a finite number of
    direct implications. As the loops are of order $\log{N}$ these
    implications never lead to a contradiction. At the spinodal point
    $\theta_s$, fixing one more variable generates an extensive avalanche of
    direct implications. Small (order $1/N$) errors in the previously
    used BP marginals may thus lead to a contradiction. This indeed happens in
    almost all the runs we have done. For more detailed discussion see
    \cite{MontanariRicci07}.}
\item{No more free variables -- The second reason for the failure is specific
    to the locked problems, more precisely to the problems where 
$\phi_0=\phi_1=1$ is a 
    solution of eqs.~(\ref{eq:qtophi}-\ref{eq:qs}). In these cases it may happen that the function
    $\Phi(\theta)\to 1$ at some $\theta_1<1$ (e.g. $\theta_1\approx 0.73$ at $L=4$
    for the  1-or-3-in-5 SAT problem). In other words if we reveal a fraction
    $\theta>\theta_1$ of variables from a random solution, the reduced problem
    will be compatible with only that given solution. Again, if there has been a little error in
    the previously fixed variables, the BP uniform decimation ends up in a
    contradiction. If on the contrary the function $\Phi(\theta)$ reaches the
    value 1 only for $\theta=1$ then the residual entropy is positive and
    there might be at each step some space to correct previous small errors,
    demonstrated on a non-locked problem in Fig.~\ref{Fig:BP_dec}. }
\end{itemize}

These two reasons of failure of the BP uniform decimation seem to be of quite
different nature. But they have one property in common. As the point of failure is approached we observe that almost all the variables which are being fixed were already implied. The same sign of failure can be observed also in the maximal BP decimation. In Fig.~\ref{Fig:BP_dec} we compare the two procedures. On the $x$-axes we plot the number of variables which could have taken both the values just before they were fixed. On the $y$-axes the number of variables which could take only one value before they were fixed plus the number of implied variables. The failure of both the versions of the BP decimation algorithms is then related to the divergence of the derivative of the function $y(x)$.

\subsection{The BP reinforcement algorithm}

BP reinforcement is currently the most efficient way of using the BP equations in a solver. It was originally introduced in
\cite{ChavasFurtlehner05}, and has also been used in \cite{BraunsteinZecchina06,DallAstaRamezanpour08}. The main idea is to add an `external bias'
$\mu_{s_i}^i$ which biases the variable $i$ in the direction of the marginal
probability computed from the BP messages. This modifies BP eq.~(\ref{BP2})
to 
\begin{subequations}
\bea
     \psi_{s_i}^{a\to i} &=& \frac{1}{Z^{a\to i}} \sum_{\{s_j\}} \delta_{A_{s_i+\sum_{j} s_j},1}
  \prod_{j\in \partial a-i} \chi_{s_j}^{j\to a}\, , \label{BP1_r}  \\
     \chi_{s_i}^{i\to a} &=& \frac{1}{Z^{i\to a}} \, \mu^i_{s_i} \prod_{b\in \partial i-a} \psi_{s_i}^{b\to i}\, , \label{BP2_r} 
\eea
\end{subequations}
We remind that the belief on variable $i$ (the BP estimate of its marginal) $ \chi_{s_i}^{i}$, without taking into account the bias $\mu$, is given by eq.~(\ref{marginal}). 

We tried several implementations of how the external bias $\mu_{s_i}^i$ is updated and found the
 best  performance for the following one
\begin{subequations}
\bea
   \mu^i_1 = \pi,  \quad  \mu^i_0 = 1-\pi,  \quad {\rm if} \quad \chi^i_{0}>\chi^i_{1} \, , \label{rein1}\\
   \mu^i_1 = 1-\pi,  \quad  \mu^i_0 = \pi,  \quad {\rm if} \quad \chi^i_{0}\le \chi^i_{1} \, , \label{rein2}
\eea
\end{subequations}
where $0\le \pi\le 1/2$ is a parameter which needs to be optimized. In the iterative update of the BP reinforcement, the external bias is not updated at every BP iteration, but only with probability  
\be
             p(t) = 1-(1+t)^{-\gamma} \, .
\ee
where $t$ is the time step, and $\gamma$ a parameter to be optimized. The pseudocode of the algorithm is then as follows. 
\begin{codebox}
\Procname{$\proc{BP-Reinforcement}(T,\gamma,\pi)$}
\li Initialize $\mu^i_{s_i}$ and $\psi_{s_i}^{a\to i}$ randomly;
\li $t\gets 0$;
\li Compute the current configuration $r_i={\rm argmax}_{s_i} \mu^i_{s_i}$;
\li \Repeat Make one sweep of the BP iterations (\ref{BP1_r}-\ref{BP2_r}); \label{line_BP}
\li         update every bias $\mu_{s_i}^i$ with probability $p(T)$ according to (\ref{rein1}-\ref{rein2});
\li         Update $r_i={\rm argmax}_{s_i} \mu^i_{s_i}$;
\li         $t\gets t+1$;
\li \Until  $\{r\}$ is a solution or $t>T$; 
\end{codebox}

This algorithm depends on two empirical parameters, $\gamma$ and $\mu$. We generally  use $\gamma=0.1$.
The optimization of the bias strength $\pi$ is crucial. Empirically we observed three different regimes:
\begin{itemize}
   \item[(a)]{$\pi_{\rm BP-like}<\pi<0.5 $: When the bias is weak, \proc{BP-Reinforcement} converges very fast to a BP-like fixed point, the values of the local fields do not point towards any solution. On contrary many constraints are violated by the final configuration $\{r_i\}$.}
   \item[(b)]{$\pi_{\rm conv}< \pi < \pi_{\rm BP-like} $: \proc{BP-Reinforcement} converges to a solution $\{r_i\}$.}
   \item[(c)]{$0 < \pi < \pi_{\rm conv}$: When the bias is too strong, \proc{BP-Reinforcement} does not converge. And many constraints are violated by the configuration $\{r_i\}$ which is reached after $T_{\rm max}$ steps.}
\end{itemize}
When the constraint density in the CSP is large the regime (b) disappears and
$\pi_{\rm conv}= \pi_{\rm BP-like}$. Clearly, the  goal is to find
$\pi_{\rm conv}<\pi<\pi_{\rm BP-like}$. The point $\pi_{\rm BP-like}$ is very
easy to find, because for larger $\pi$ the convergence of
\proc{BP-Reinforcement} to a BP-like fixed point takes place in just a few
sweeps. Thus in all the runs we chose $\pi$ to be just below $\pi_{\rm
  BP-like}$. The value of $\pi$ chosen in this way does not seem to
depend on the size of the system, but it depends slightly on the
constraint density.

\begin{figure}[!ht]
  \resizebox{\linewidth}{!}{
  \includegraphics{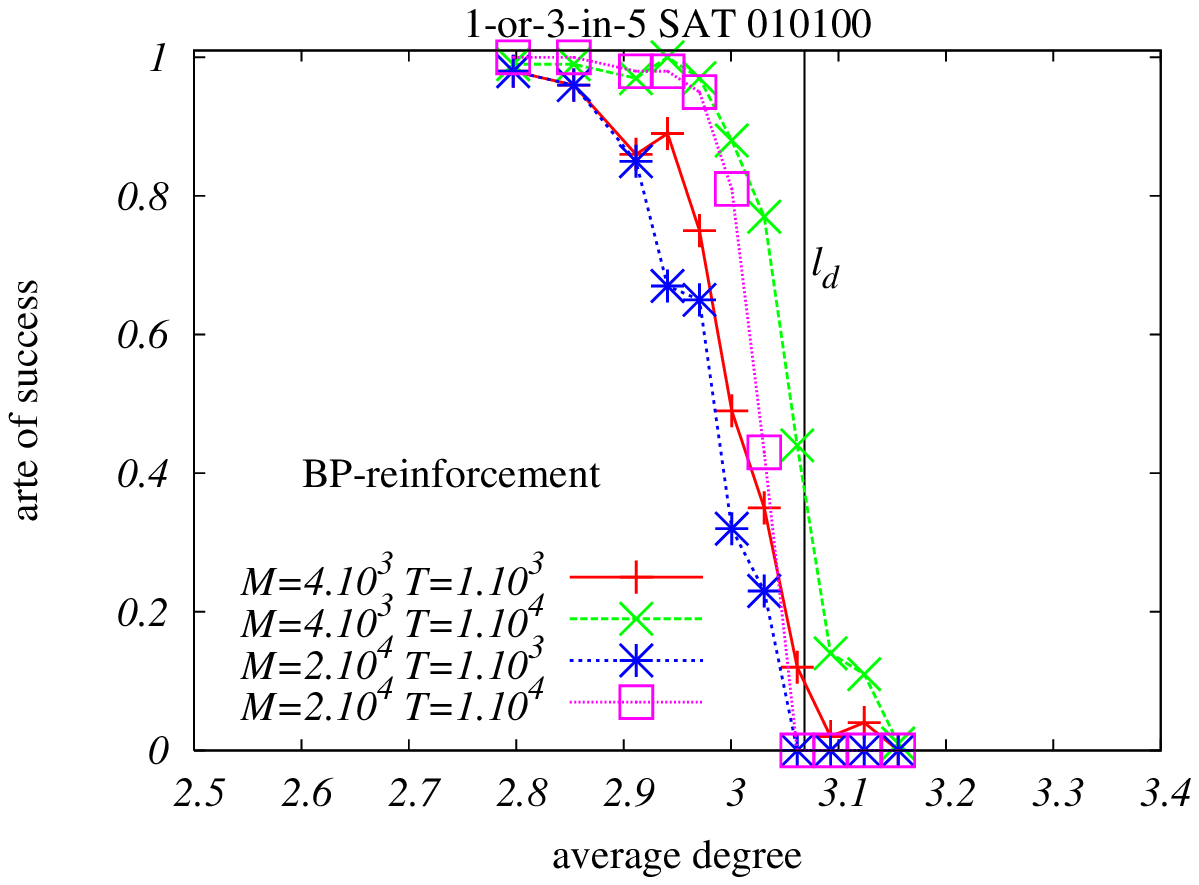}
  \includegraphics{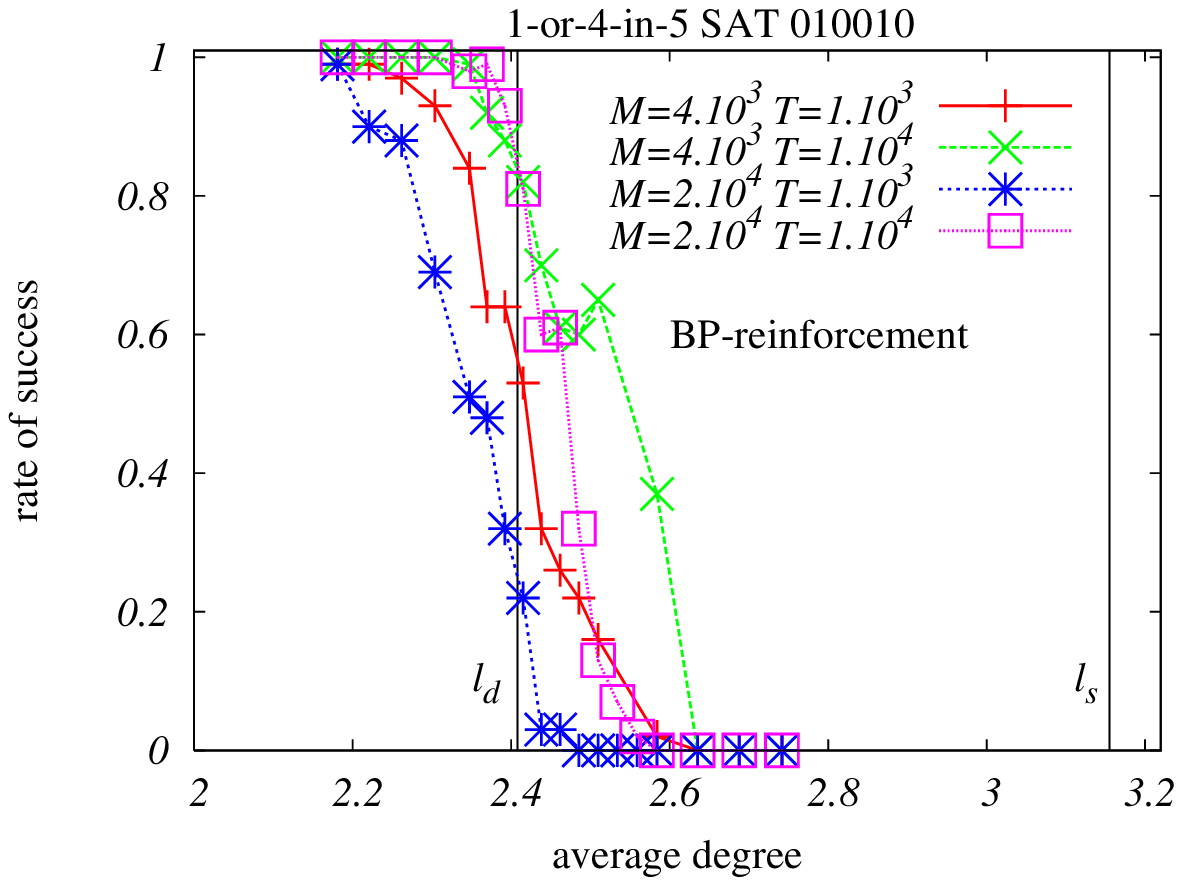}
  }
  \caption{\label{Fig:BP_locked} Performance of the BP reinforcement on two of the locked occupation problems. 
Probability of success versus average connectivity. Left: $A=010100$, the optimal parameters: $\gamma=0.1$, $\pi=0.28$ for $2.79 \le \overline l \le 2.95$, $\pi=0.30$ for $2.97 \le \overline l \le 3.13$, $\pi=0.31$ for $c = 3.15$. Right: $A=010010$ with $\gamma=0.1$, $\pi=0.34$. The different curves are for two different system sizes and two different maximal running times. The algorithm performs well only up to a connectivity close to the clustering transition ($l_d=3.07$ resp. $l_d=2.41$ to be compared with the satisfiability threshold $l_s=4.72$ resp. $l_s=3.16$). Qualitatively similar result were observed for all the other locked occupation problems we studied.}
\end{figure}

We tested the \proc{BP-Reinforcement} algorithm on the locked occupation CSPs,
the results are shown in Fig.~\ref{Fig:BP_locked}. The fraction of successful
runs on different system sizes and for different maximal running times is
plotted as a function of the mean variable connectivity. Our data suggest that the
algorithm is successful only in the liquid phase, and fails in the clustered
(that is also frozen) region. Similar results
can be obtained with other algorithms; for
instance the performance of stochastic local search was reported in
\cite{ZdeborovaMezard08}.

The clustered phase is thus extremely hard and instances of the locked
problems can serve as benchmarks for new solvers. In fact, some of the hardest
benchmarks of the K-satisfiability problem are based on a well known LOP,  XOR-SAT
(with some additional non-linear function nodes which rule out the Gaussian
elimination solvers) \cite{HaanpaaJarvisalo06,BarthelHartmann02}.

\begin{figure}[!ht]
  \resizebox{\linewidth}{!}{
  \includegraphics{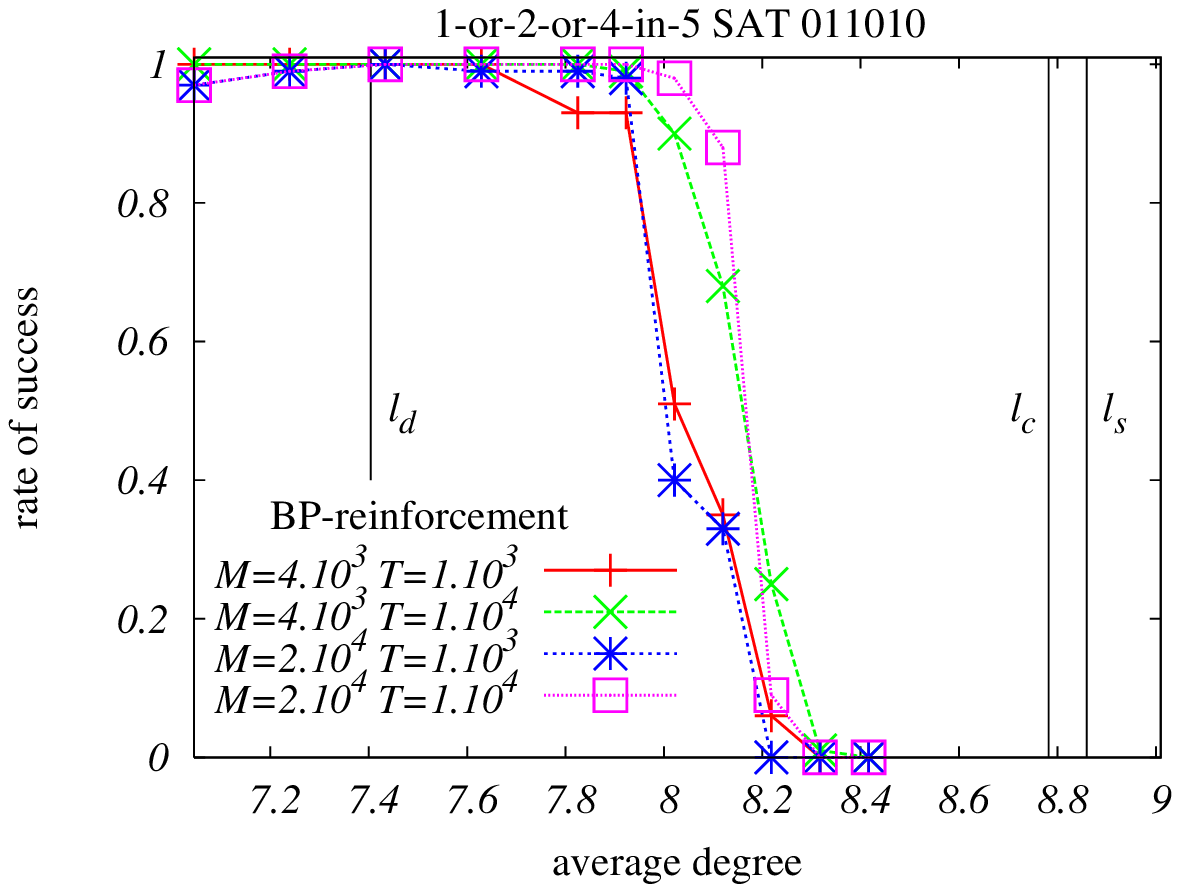}
  \includegraphics{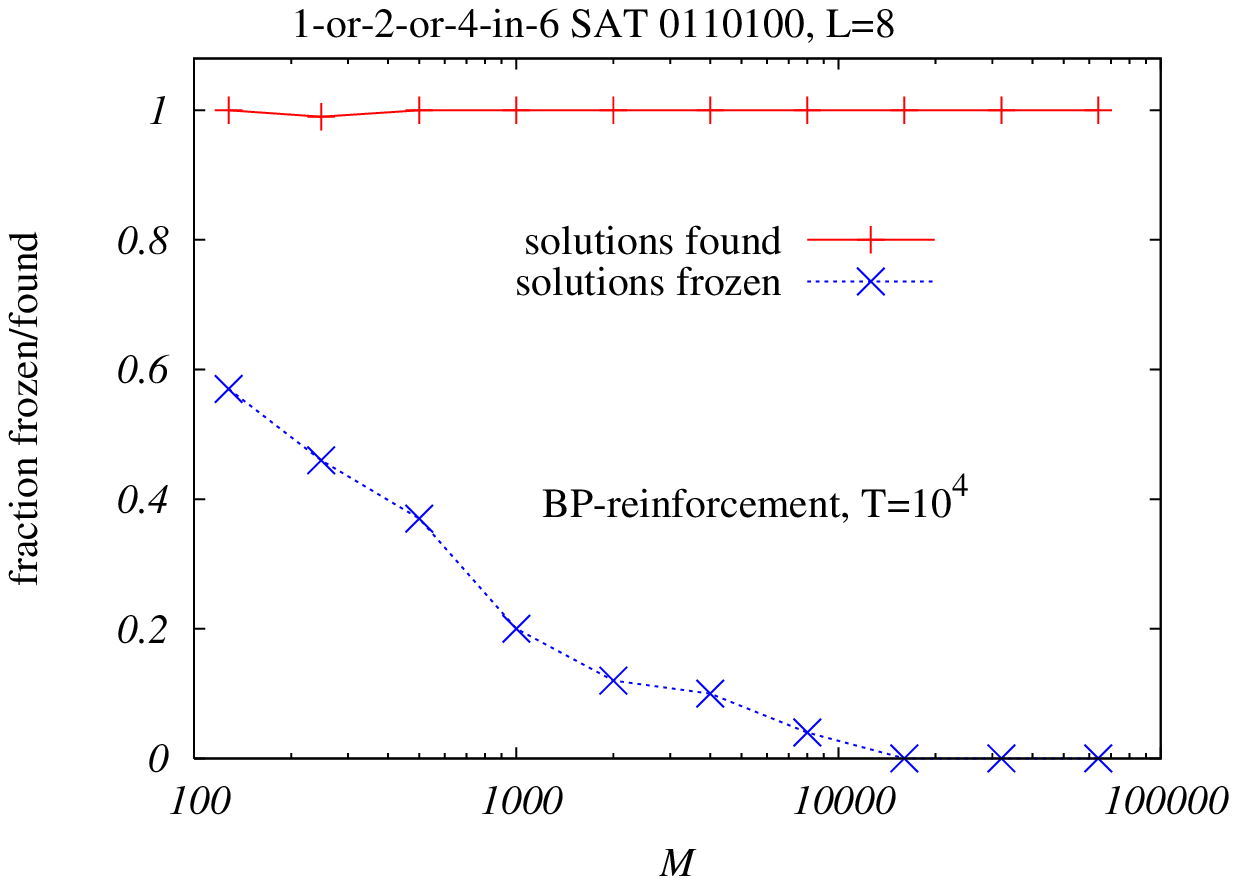}}
  \caption{\label{Fig:BP_non} Left: Performance of the BP reinforcement on one of the non-locked problems, $A=010110$. Parameters: $\gamma=0.1$; $\pi=0.40$ for $7.0 \le \overline l \le 7.8$, $\pi=0.42$ for $7.9 \le \overline l \le 8.0$, $\pi=0.44$ for $8.1 \le \overline l \le 8.4$. The implementation of the algorithm is the same as for the locked problems in Fig.~\ref{Fig:BP_locked}. Here solutions are found up to about a half of the clustered region, $l_d=7.40$. The condensation $l_c=8.78$ and the satisfiability $l_s=8.86$ transition are also marked.    
Right: The $A=0110100$ at regular graphs of $L=8$ is in the rigid phase, that is almost all solutions belong to frozen clusters. Yet the BP reinforcement ($\gamma=0.1$, $\pi=0.36$) finds a solution almost surely (after 3 restarts) -- the red curve. The blue curve gives a fraction of how many of the solutions found belonged to a frozen cluster. We see that asymptotically we never find the frozen solutions.}
\end{figure}

In the non-locked problems the very same implementation of the BP reinforcement is able to find solutions inside the clustered region, Fig.~\ref{Fig:BP_non} left shows the performance for $A=011010$. This is in qualitative agreement with results for the K-SAT  \cite{ChavasFurtlehner05}, coloring \cite{KrzakalaMontanari06,ZdeborovaKrzakala07} or bicoloring problems \cite{DallAstaRamezanpour08}. 

It is not known how one can characterize from a geometrical point of view the
connectivity threshold where BP reinforcement algorithms stop to be efficient
in the non-locked problems. It has been found in \cite{DallAstaRamezanpour08}
that even the rigid phase where almost all solutions are frozen may be
algorithmically easy. Fig.~\ref{Fig:BP_non} confirms this statement for the
problem $A=0110100$ on the regular ensemble with $L=8$. The ratio of success
of the BP reinforcement (with 3 restarts) is close to one, and basically
independent of system size, while it can be seen from (\ref{eq:mus}) that this
problem is in the rigid phase. On the other hand the fraction of found
solutions which are frozen (have a nontrivial whitening core
\cite{BraunsteinZecchina04,ArdeliusZdeborova08}) goes to zero as the system
size is growing, in agreement with the results of \cite{DallAstaRamezanpour08}. Thus the question of where is the easy/hard threshold in the non-locked problems remains open.

\section{Conclusion}
\label{sec:con}

We studied the class of occupation CSPs on which we illustrated the difference
between locked and non-locked CSPs. The point-like nature of clusters in LOPs is
responsible for all of these differences. Our finding may be summarized as:
"Locked problems are extremely simple and extremely hard." The simplicity
comes at the level of the phase diagram, which can be computed by the cavity
method much more easily that in the general CSP. In certain cases some
non-trivial quantities are probably amenable to a rigorous study along the
lines that we sketched -- as for example the satisfiability threshold in the
balanced locked problems. The hardness is algorithmic, some algorithms -- as
the BP decimation -- fail completely, and even the best known algorithms are
not able to find solutions in the clustered phase of the locked problems. Their
simple description and algorithmic hardness makes the locked problems 
challenging for developments of new algorithms as well as for better theoretical
understanding on the origin of hardness.

There are several clear directions in which this work should be extended. The
planted ensembles of LOPs should be studied in order to provide hard
benchmarks where the existence of a solution would be guaranteed. On the
mathematical side the rigorous proof of the second moment method giving the
satisfiability threshold in the balanced locked problems should be worked out.
One may investigate if the location of the clustering threshold can be proven
rigorously using the small noise reconstruction in the lines of \cite{Sly08},
or bounding the weight enumerator function in the lines of
\cite{MoraMezard06}. Also it will be interesting to study (at least
numerically) the dynamics at finite temperature, as it might provide further
insight into the connection between the dynamics of algorithms and structure
of solutions. Finally the distance properties between solutions in locked CSPs
makes them interesting candidates for the development of nonlinear error
correcting codes or compression schemes.

\section*{Acknowledgment}

We thank Florent Krzakala, Cris Moore, Thierry Mora and Guilhem Semerjian for very useful discussions.

\bibliography{myentries}

\end{document}